# Projection Operator

# The Mori-Zwanzig method of projection operators: Generalized Langevin equation.

(Lecture Notes)

**N.F. Fatkullin**

A fairly brief and complete presentation of the Zwanzig - Mori projection operator technique is given.

**Table of Contents**

**Chapter 1. General Information.**

## Chapter 1: General Information.

### 1.1 General statements of the Liouville formalism.

#### 1.1.1 The classical Liouville space.

Consider a classical system of N particles. The set of 6N generalized coordinates and momenta $\Gamma \equiv \{Q_1, Q_2, \ldots, Q_{3N,} P_1, P_2, \ldots, P_{3N}\}$defines the dynamical state of the system in the mechanical description and can be regarded as a point of a 6N-dimensional phase space F. Any physical property of our system corresponds to some function of generalized coordinates and momenta $A(\Gamma) \equiv A(Q_1, \ldots, Q_{3N,} P_1, \ldots, P_{3N})$, i.e., a function defined on the phase space **F**. Mathematically, in many respects it is convenient to consider the functions $A(\Gamma)$ , sometimes called observables, as complex-valued. Observables are usually required to be infinitely differentiable almost everywhere on **F** .

Let us denote the set of all observables by **L**. The set **L** can be considered as an infinite-dimensional linear space over the field of complex numbers. Indeed, if $A(\Gamma),B(\Gamma) \in \mathbf{L}$, then $A(\Gamma)+B(\Gamma) \in \mathbf{L}$. If $\alpha$ is a complex number and $A(\Gamma) \in \mathbf{L}$ , then $\alpha A(\Gamma) \in \mathbf{L}$ . One can specify any number of linearly independent functions, such as sets of polynomials of different degrees, so the dimension of L is infinitely large. Moreover, since the product $A(\Gamma) \bullet B(\Gamma) \in \mathbf{L}$ , if $A(\Gamma),B(\Gamma) \in \mathbf{L}$ , then the space **L** is an infinite-dimensional commutative algebra over the field of complex numbers.

Thus, physical quantities $A(\Gamma)$ can be considered as vectors of space **L**. The operation of scalar product of two vectors can be defined in many ways on the space **L**. The most natural way from the point of view of statistical physics is the following. The results of any experimental measurements of the properties of macroscopic systems give information about some correlation functions depending on the type of a particular experiment. Among all types of correlation functions, the simplest are the equilibrium binary correlation functions of two physical quantities $A^*(\Gamma)$ and $B(\Gamma)$ defined by the relation:

$$\langle A(\Gamma)|B(\Gamma)\rangle \equiv \left\langle A^*(\Gamma)B(\Gamma)\right\rangle_{eq} = \int d\Gamma A^*(\Gamma)B(\Gamma)\rho_{eq}(\Gamma) \qquad (1.1)$$

where $\rho_{eq}(\Gamma)=\frac{1}{Z}\exp\{-\beta H(\Gamma)\}$ is equilibrium Gibbs distribution function, $\beta=(kT)^{-1}$ is incerse temperature, $k$ is Boltzmann constant, T is absolute temperature, $Z=\int d\Gamma \exp\{-\beta H(\Gamma)\}$ – classical statistical sum (or statistical integral), $H(\Gamma)$ –Hamiltonian of the considered system, assumed to be a given real function from the space **L**, $A^*(\Gamma)$ – function complex conjugate to $A(\Gamma)$.

Note that in all physically interesting situations the properties of the Hamiltonian are such that the relation (1.1) can be considered defined on all pairs of vectors from L without essential restriction of the space **L** itself.

The correlation function $\left\langle A^*(\Gamma)B(\Gamma)\right\rangle_{eq}$ can be considered as the scalar product $\langle A|B\rangle$ of two vectors $A(\Gamma),B(\Gamma)\in$ **L**.

Indeed, it is easy to check the validity of the following relations defining the scalar product on complex spaces:

1. $\left(\langle A|B\rangle\right)^{*} = \langle B|A\rangle$
2. *Если* $B = \alpha_1 B_1 + \alpha_2 B_2$ , *то* $\langle A|B\rangle = \alpha_1 \langle A|B_1\rangle + \alpha_2 \langle A|B_2\rangle$ (1.2)
3. $\forall A(\Gamma) \neq 0 \;\; \langle A|A\rangle > 0.$

*Exercises.1.1*

*1. Let the Hamiltonian of the system have the standard*

$$H(\Gamma) = \sum_i \frac{\vec{p}_i^{\,2}}{2m} + \sum_{i<j} U(\vec{r}_{i\,j}),$$

*where* $\vec{p}_i$ *– momentum of i-th particle ,* $U(\vec{r}_{i\,j})$ *is the potential energy of interaction of particles with numbers i and j. A "test" particle is one of the particles in the system. Let's* $p^{\alpha}$ *- momentum projection of the "test" particle on the axis* $\alpha$*. Show that*

$$\langle p^{\alpha}|p^{\alpha}\rangle = mkT$$

$$\langle \vec{p}|\vec{p}\rangle = 3mkT,$$

$$\langle \vec{p}|\vec{p}\rangle \equiv \langle p_x|p_x\rangle + \langle p_y|p_y\rangle + \langle p_z|p_z\rangle$$

*2. Let* $\vec{r}$ *radius - vector of the "test" particle.Show, that* $\langle \vec{r}|\vec{r}\rangle = R^2$, $R$ *is the radius of inertia of the whole system (it is assumed that the system consists of identical particles and the center of mass coincides with the origin;* $\langle \vec{r}|\vec{r}\rangle = \langle x|x\rangle + \langle y|y\rangle + \langle z|z\rangle$.

*3. Show, that*

$$\langle x^{\alpha}|x^{\beta}\rangle = \frac{1}{3}R^2\delta_{\alpha\,\beta}$$

$$\langle p^{\alpha}|p^{\beta}\rangle = mkT\delta_{\alpha\,\beta}$$

$$\langle x^{\alpha}|p^{\beta}\rangle = 0$$

4. $\left\langle \frac{m\upsilon^2}{2} \middle| \frac{m\upsilon^2}{2} \right\rangle = \frac{15}{4}(mkT)^2$

5. $\left\langle \vec{p} \middle| \frac{m\upsilon^2}{2} \right\rangle = 0$

The space of functions **L** in which the scalar product is defined by the relation (1.1) is called Liouville space. From the mathematical point of view, the Liouville space can be considered as an example of a commutative normalized algebra possessing all the properties of a Hilbert space.

Any function from L is a vector of Liouville space. By analogy with the quantum-mechanical terminology and notations proposed by P. Dirac, we can define "bra" – vectors $\langle A| \equiv A^*(\Gamma)$ and "cket" are vectors $|B\rangle \equiv B(\Gamma)$. The scalar product can be viewed as the product of a "bra"-vector by a "cket"-vector $\langle A|B\rangle$.

### 1.1.2 Liouville superoperator (Liouvillian)

The state of the system in the statistical-mechanical description is given by the distribution function $\rho(\Gamma,t)$ on the phase space F. The function $\rho(\Gamma,t)$ depends parametrically on time t and determines the probability density at time t of the system under study to have generalized coordinates and impulses corresponding to the point $\Gamma$ of the phase space F. The state of the system evolves with time, satisfying the Liouville equation:

$$\frac{\partial}{\partial t}\rho(\Gamma,t) = \{H\, ;\rho\} \tag{1.3}$$

where

$$\{A(\Gamma);B(\Gamma)\} \equiv \sum_{i=1}^{3N}\left(\frac{\partial A}{\partial Q_i}\frac{\partial B}{\partial P_i} - \frac{\partial A}{\partial P_i}\frac{\partial B}{\partial Q_i}\right)$$

- Poisson bracket.

Distribution function $\rho(\Gamma,t)\in\mathbf{L}$. Consider an arbitrary function $A(\Gamma)\in$ **L**. The Poisson bracket operation is linear on each argument, so the operator $\hat{L}$, defined on **L**, as

$$\hat{L}\equiv i\{H;\ldots\}, \tag{1.4a}$$

is a linear. Equation (1.3) can be given the form of the Schrödinger equation by multiplying by the imaginary unit i both parts of the equation (1.3) :

$$i\frac{\partial}{\partial t}|\rho(t)\rangle=\hat{L}|\rho(t)\rangle \tag{1.4}$$

The linear operator $\hat{L}$ is called the Liouville superoperator or Liouvillian of the system. Note that the term "superoperator" is used for uniformity of terminology at generalization to quantum-mechanical systems. In this case the Liouville space appears as a set of operators acting on the Hilbert space defined by wave functions of the system. Then $\hat{L}$ is an operator acting on the operator space. The prefix "super" is used to emphasize this circumstance. The linear operator $\hat{L}$ is an Hermite, or self-adjoint, operator acting in Liouville space.

Indeed, consider a scalar product

$$\left\langle A\middle|\hat{L}B\right\rangle\equiv\int d\Gamma A^{*}(\Gamma)i\{H;B\}\rho_{eq}(\Gamma) \tag{1.5}$$

Using the definition for the Poisson bracket (explanation of relation (1.3)) and the fact that $\rho_{eq}$ depends only on the Hamiltonian of the system, it is easy to verify the validity of the relation:

$$\{H(\Gamma);B(\Gamma)\}\rho_{eq}(H(\Gamma))=\{H;B\rho_{eq}\} \tag{1.6}$$

Substituting this equality into the right part of the relation (1.5), and taking into account that $\rho_{eq}(\Gamma)$ decreases fast enough at the phase space boundaries, we obtain after integration by parts

$$\left\langle A\middle|\hat{L}B\right\rangle=-\int d\Gamma i\{H;A^{*}\}B\rho_{eq}(\Gamma)\ . \tag{1.7}$$

The Hamiltonian of the system *H(Γ)* is always a real-valued function. This allows us to write the right part of equality (1.7) in the form:

$$\left\langle A \middle| \hat{L}B \right\rangle = \int d\Gamma \left( i\{H\, ;A\} \right)^* B\rho_{eq}(\Gamma) \qquad (1.8)$$

Recall that the Hermite conjugate operator $\hat{L}^+$ to an operator $\hat{L}$ must satisfy the equality $\hat{L}^+$

$$\left\langle A\hat{L}^+ \middle| B \right\rangle = \left\langle A \middle| \hat{L}B \right\rangle \qquad (1.9)$$

for any functions *A(Γ)* and *B(Γ)* from **L**.

From this definition and equality (1.8) we see that $\hat{L} = \hat{L}^+$ , i.e., the superoperator $\hat{L}$ is Hermite-conjugate to itself, i.e., Hermite (in the physical literature, as a rule, no distinction is made between Hermite and self-adjoint operators).

### 1.1.3. Heisenberg and Schrödinger representations.

The formalism of the classical Liouville space, as already mentioned, is based on the formal analogy between classical statistical mechanics and quantum mechanics. As is well known, quantum mechanics emphasizes two basic representations: Schrödinger and Heisenberg, analogs of which are also available within the classical Liouville space formalism. In the Schrödinger representation, the state of the system $\rho(\Gamma,t)$, in general, depends on time, satisfying the Liouville equation (1.4). If there are no external influences on the system, its Hamiltonian *H(Γ)* does not depend on time. Therefore, it is independent of time and the Liouville superoperator $\hat{L}$, which allows the solution of equation (1.4) to be represented as:

$$\rho(\Gamma,t) = \exp\left\{ -i\hat{L}t \right\} \rho_0(\Gamma) \qquad (1.10)$$

where

$$\exp\left\{-i\hat{L}t\right\} \equiv \sum_{n=0}^{\infty} \frac{(-it)^n}{n!} \hat{L}^n \qquad (1.10a)$$

$\rho_0(\Gamma)$ - distribution function of the system at time *t=0*. Exponential operator (superoperator) $\exp\left\{-i\hat{L}t\right\}$ describes the change with time of the state of the system and is called a superoperator of evolution or propagator.

*Exercise.1.2.*

*Make sure that the equilibrium state of the system does not change under the action of the* $\exp\left\{-i\hat{L}t\right\}$.

The functions *A(Γ)* on the phase space describing physical quantities do not change with time. The experimentally observed properties of a physical system are the mean values of *A(Γ)* under the distribution $\rho(\Gamma,t)$:

$$\langle A(t)\rangle = \int d\Gamma A(\Gamma)\rho(\Gamma,t) \qquad (1.11)$$

If the system is not in equilibrium state, $\langle A(t)\rangle$ depends on time due to the evolution of the system state, i.e. $\rho(\Gamma,t)$.

Using the relation (1.10), let us rewrite the equality (1.11) in the form:

$$\langle A(t)\rangle = \int d\Gamma A(\Gamma)\exp\left\{-i\hat{L}t\right\}\rho_0(\Gamma) \qquad (1.12)$$

Let us now decompose the superoperator of evolution into a taylor series according to the relation (1.10a). Then, we will reintegrate each term of the expansion by parts so that the action of the operator $\hat{L}^n$ is transferred from $\hat{L}^n$ to *A(Γ)*. As a result, we get:

$$\langle A(t)\rangle = \int d\Gamma \rho_0(\Gamma)\exp\left\{+i\hat{L}t\right\} A(\Gamma) \qquad (1.13)$$

The physical quantity

$$A(t) \equiv A(\Gamma,t) \equiv \exp\left\{+i\hat{L}t\right\} A(\Gamma) \qquad (1.14)$$

is by definition called A in the Heisenberg representation.

Hereinafter, for the sake of brevity, the following designations will be used

$$A(t) \equiv A(\Gamma, t) \tag{1.15a}$$

$$A(t{=}0) \equiv A \equiv A(\Gamma) \tag{1.15b}$$

It is easy to see that $A(t)$ satisfies the equation

$$\frac{d}{dt}A(t) = i\hat{L}_H A(t) \tag{1.16}$$

The relation (1.14) is a formal solution of equation (1.16), the relation (1.13) is equivalent to the following:

$$\langle A(t)\rangle = \int d\Gamma A(t)\rho_0(\Gamma) \tag{1.17}$$

Thus, in the Heisenberg representation, the state of the system $\rho_0(\Gamma)$ does not change with time, and the functions describing physical quantities evolve with time in accordance with relations (1.14) and (1.16). By analogy with quantum mechanics one can define the Dirac representation, or the interaction representation. However, we will not need it further, so its detailed consideration is presented to the reader as an independent exercise.

## 1.2. Derivation of kinetic equations.

### 1.2.1 Projection operators.

The scalar product of two physical quantities in space **L** is defined by the relation (1.1). This allows us to define projection operators on a physical quantity $|A\rangle = A(\Gamma)$ and on linear subspaces stretched on certain physical quantities $|A_1\rangle$, ..., $|A_n\rangle$.

Consider a certain physical quantity $|A\rangle = A(\Gamma)$. An algebraic construction

$$\hat{P} = |A\rangle \frac{1}{\langle A|A\rangle}\langle A| \tag{1.18}$$

can be interpreted as a projection operator on the physical quantity $A(\Gamma)$ if we define its action on an arbitrary function $|B\rangle = B(\Gamma)$ as follows:

$$\hat{P}|B\rangle = |A\rangle \frac{\langle A|B\rangle}{\langle A|A\rangle} = A(\Gamma)\frac{\int d\Gamma A^*(\Gamma)B(\Gamma)\rho_{eq}(\Gamma)}{\int d\Gamma A^*(\Gamma)A(\Gamma)\rho_{eq}(\Gamma)} \tag{1.19}$$

Geometrically, a vector $\hat{P}|B\rangle$ can be viewed as the projection of a vector in Liouville space onto a vector $|A\rangle$. It is easy to see that the operator

$$\hat{P}^2 = \hat{P},\ \hat{P}|A\rangle = |A\rangle \tag{1.20}$$

*Exercise.*

*Prove that the operator* $\hat{P}$ *is Hermite operator, i.e.*

$$\left\langle A\middle|\hat{P}B\right\rangle = \left\langle A\hat{P}\middle|B\right\rangle \tag{1.21}$$

Operators having the above properties (relations (1.20) and (1.21)) are called projection operators. The operator

$$\hat{Q} = 1 - \hat{P} \tag{1.22}$$

also has all these properties, it is a projection operator on the orthogonal addition to the one-dimensional subspace defined by $|A\rangle$.

Let us consider a more general case when we are interested in several linearly independent quantities $A_1(\Gamma)$, $A_2(\Gamma)$, ..., $A_n(\Gamma)$. The set of all linear combinations of these quantities forms a linear subspace $\mathbf{L_n}$ of dimension n in Liouville space. Let us construct a projection operator on this subspace, which is sometimes simplified as “the projection operator on the values $A_1$, ..., $A_n$”. Consider a formal sum

$$\hat{P} = \sum_{k,l} |A_k\rangle \alpha_{k\,l} \langle A_l| \tag{1.23}$$

where $\alpha_{k\,l}$ - some given numbers forming a matrix $\hat{\alpha}$ of size $n\times n$, $k,l=1,2,...n$.

For simplicity, we will not supply the projection operator on $\mathbf{L_n}$ with additional indices and denote by the same letter as defined above. This will not lead to confusion anywhere in the future. This sum can be considered as a linear operator translating any vector $|B\rangle \in \mathbf{L}$ into some vector from $\mathbf{L_n}$ by the rule:

$$\hat{P}|B\rangle \equiv \sum_{k,l} |A_k\rangle \alpha_{k\,l} \langle A_l|B\rangle \tag{1.24}$$

Let's pick the numbers $\alpha_{k\,l}$ so that $\hat{P}$ there is a projection operator.

$$\hat{P}^2 = \hat{P} \tag{1.25}$$

Substituting the ratio (1.23) into both parts of the equality (1.25), we obtain:

$$\sum_{\substack{k,m \\ m',l}} |A_k\rangle \alpha_{k\,m} \langle A_m|A_{m'}\rangle \alpha_{m'\,l} \langle A_l| = \sum_{k,l} |A_k\rangle \alpha_{k\,l} \langle A_l| \quad . \tag{1.26}$$

The set of numbers $\langle A_m|A_{m'}\rangle$ forms a matrix $\langle A \otimes A\rangle$ of size $n \times n$ , which due to the linear independence of vectors $A_1, \ldots, A_n$ is nondegenerate, i.e. $\det\langle \hat{A} \otimes \hat{A}\rangle \neq 0$ and there exists an inverse matrix $\langle A \otimes A\rangle^{-1}$-1. It is easy to see that the relation (1.26) holds if the equality is satisfied:

$$\sum_{m,m'} \alpha_{k\,m} \langle A_m|A_{m'}\rangle \alpha_{m'\,l} = \alpha_{k\,l} \tag{1.27a}$$

or in matrix notation:

$$\hat{\alpha} \langle A \otimes A\rangle \hat{\alpha} = \hat{\alpha} \ . \tag{1.27b}$$

The solution of the matrix equation (1.27b) is

$$\hat{\alpha} = \langle A \otimes A\rangle^{-1} \tag{1.28}$$

Let's define the operator as follows:

$$\hat{P} = \sum_{k,l} |A_k\rangle \left[\frac{1}{\langle A \otimes A\rangle}\right]_{k\,l} \langle A_l| \tag{1.29}$$

where $\left[\frac{1}{\langle A \otimes A\rangle}\right]_{k\,l}$ denote matrix elements of the matrix $\langle \hat{A} \otimes \hat{A}\rangle^{-1}$.

*Exercise*

*1. Prove that for any vector* $A \in \mathbf{L_n}$ $\hat{P}\,|A\rangle = |A\rangle$.

*2. Prove that the operator* $\hat{P}$ *is Hermitian.*

The operator , defined by relation (1.29), is thus a projection operator on a subspace $\mathbf{L_n}$ stretched over an arbitrary linearly independent set of physical quantities $A_1(\Gamma), A_2(\Gamma), \ldots, A_n(\Gamma))$.

If the quantities $A_k$ are mutually orthogonal, i.e. $\langle A_k \,|\, A_l \rangle = \delta_{k\,l} \langle A_k \,|\, A_k \rangle$ , then the projection operator $\hat{P}$ is the sum of projection operators on individual physical quantities $A_k$:

$$\hat{P} = \sum_{k=1,\ldots n} |A_k\rangle \frac{1}{\langle A_k \,|\, A_k \rangle} \langle A_k| \qquad (1.30)$$

The projection operator on the orthogonal completion of $\mathbf{L_n}$ is given by the relation

$$\hat{Q} = 1 - \hat{P} \quad . \qquad (1.31)$$

### 1.2.2 Kubo operator identities.

In the following we will need a relation called the Kubo operator identity:

$$e^{i(\hat{A}+\hat{B})t} = e^{i\hat{A}t} + \int_0^t d\tau e^{i(t-\tau)(\hat{A}+\hat{B})} i\hat{B} e^{i\hat{A}\tau} \qquad (1.32)$$

For the proof we introduce an additional operator $\hat{J}(t)$, defined by the equality:

$$\hat{J}(t) = e^{it(\hat{A}+\hat{B})} e^{-it\hat{A}} \,.. \qquad (1.33)$$

Differentiating equality (1.33) by the parameter t, we obtain:

$$\frac{d}{dt}\hat{J}(t) = e^{it(\hat{A}+\hat{B})} i\hat{B} e^{-it\hat{A}} \qquad (1.34)$$

Let us integrate the last equality over time:

$$J(t) = I + \int_0^t d\tau \exp\left\{i\tau(\hat{A}+\hat{B})\right\} i\hat{B} \exp\left\{-i\tau\hat{A}\right\}, \qquad (1.35)$$

where I is a unit operator, which is the integration constant in (1.35) (from the definition (1.33) we see that J(0)=I). Multiplying both parts of $\exp\left\{it\hat{A}\right\}$the relation (1.35) by $\exp\left\{it\hat{A}\right\}$

$$e^{i(\hat{A}+\hat{B})t} = e^{i\hat{A}t} + \int_0^t d\tau e^{i\tau(\hat{A}+\hat{B})} i\hat{B} e^{i(t-\tau)\hat{A}} \qquad (1.36)$$

*Exercise.1.3.*
*Prove the equivalence of relations (1.32) and (1.36).*

### 1.2.3. Generalized Langevin equation.

Any experimental research method provides information about a certain finite set of physical quantities or their correlation functions, determined by the research method itself. Let us denote this set of quantities by $A_1(\Gamma)$, $A_2(\Gamma)$, ..., $A_n(\Gamma)$.

For example, in methods related to magnetic resonance, this set of values is the components of the magnetic moment of the spin subsystem under study; in dielectric spectroscopy, it is the components of the electric dipole moment of the system; in neutron scattering, it is the coherent and incoherent structural factors of the system, etc.

Deriving equations describing the dynamics of these "interesting" quantities $A_1(t)$, $A_2(t)$, ..., $A_n(t)$ is one of the central problems of physical kinetics. Without losing generality, we will further assume that the values of $A_k(\Gamma)$ are defined such that their equilibrium mean values are $\langle A_k(\Gamma)\rangle_{eq} = 0$.

Let us consider one of these quantities, $A_m(t)$, in Heisenberg's representation. It satisfies the equation:

$$\frac{d}{dt}A_m(t) = i\hat{L}\exp\{i\hat{L}t\}A_m = i\exp\{i\hat{L}t\}\hat{L}A_m \ . \tag{1.37}$$

Using the identity $\hat{P}+\hat{Q}=1$ , where $\hat{P}$ is the projection operator onto $\mathbf{L_n}$ , we transform the right-hand side of (1.37) to the form:

$$\frac{d}{dt}A_m(t) = i\exp\{i\hat{L}t\}\hat{P}\hat{L}A_m + i\exp\{i\hat{L}t\}\hat{Q}\hat{L}A_m \tag{1.38}$$

Substituting expression (1.29) for the projection operator, we transform the first term on the right-hand side of relation (1.38) as follows:

$$\begin{aligned} i\exp\{i\hat{L}t\}\hat{P}\hat{L}A_m &= i\sum_{k,l}\exp\{i\hat{L}t\}\left|A_k\right\rangle\left[\frac{1}{\langle A\otimes A\rangle}\right]_{k\,l}\left\langle A_l\middle|\hat{L}A_m\right\rangle = \\ &= i\sum_k \omega_{m\,k}A_k(t) \end{aligned} \tag{1.39}$$

where

$$\omega_{m\,k} = \sum_l\left[\frac{1}{\langle A\otimes A\rangle}\right]_{k\,l}\left\langle A_l\middle|\hat{L}A_m\right\rangle \ . \tag{1.39a}$$

The system of numbers $\omega_{m\,k}$ forms the so-called frequency matrix $\hat{\omega}$ , which, according to (1.39a), is the product of

$$\hat{\omega} = \left\langle \hat{A}\otimes\hat{L}\hat{A}\right\rangle\left\langle\hat{A}\otimes\hat{A}\right\rangle^{-1} , \tag{1.40}$$

where $\left\langle \hat{A}\otimes\hat{L}\hat{A}\right\rangle$ symbolizes a matrix formed by numbers $\left\langle A_l\middle|\hat{L}A_m\right\rangle$.

*Exercise.1.4.*

*Prove that the frequency matrix can be reduced to a Hermitian matrix using a non-degenerate transformation.*

Note that the frequency matrix is completely determined by the equilibrium correlation functions $\langle A_k | A_l \rangle$ and $\langle A_k | \hat{L} A_l \rangle$. Calculating them can be a serious challenge in itself. However, in physical kinetics problems, this issue is considered to have already been solved either precisely or with some acceptable approximation.

*Exercise.1.5.*

*Let all values be characterized by a certain symmetry with respect to the time reversal operation, i.e., either even or odd. Prove that all diagonal elements of the frequency matrix are zero.*

Let us now proceed to analyze the second term on the right-hand side of equation (1.38). First, let us introduce a special

$$F_m^Q \equiv i\hat{Q}\hat{L}A_m \quad . \qquad (1.41)$$

This quantity is called the generalized stochastic force of Langevin associated with the quantity $A_m$ at the initial moment of time. The motivation for this definition will become clear from the following. For now, let us note

$$\langle A_k | F_m^Q \rangle = 0 \qquad (1.42)$$

for all $k,m=1,2,...,n$ based on construction of the values $F_m^Q$ .

*Exercise.*

*Let us choose as values $A_1(\Gamma)$, $A_2(\Gamma)$, ..., $A_n(\Gamma)$ the components $p_{x1}$, $p_{y1}$, $p_{z1}$ of the momentum of a certain "test" particle. By direct calculation, verify that $F_1^Q$, $F_2^Q$ , $F_3^Q$ are the components of the force acting on the particle under consideration from all other particles, i.e. .*

$$i\hat{Q}\hat{L}\vec{p} = -\frac{\partial}{\partial \vec{r}_1} H \, .$$

The value $\exp\{i\hat{L}t\}F_m^Q$, which is the second term in the right-hand side of formula (1.38), at an arbitrary moment of time is, in general, not orthogonal to the value $A_k$. Let us extract from it the orthogonal part, which is interpreted as the generalized stochastic Langevin force at an arbitrary moment of time. Let us extract from it the orthogonal part, which is interpreted as the generalized stochastic Langevin force at an arbitrary moment of time. For this purpose, we will represent the propagator $\exp\{i\hat{L}t\}$ in the form:

$$\exp\{i\hat{L}t\} = \exp\{i\hat{Q}\hat{L}t + i\hat{P}\hat{L}t\} \ . \qquad (1.43)$$

Next, we use the Kubo identity (1.36), assuming $\hat{A} = i\hat{Q}\hat{L}$ and $\hat{B} = i\hat{P}\hat{L}$:

$$e^{i\hat{L}t} = e^{i\hat{Q}\hat{L}t} + \int_0^t d\tau e^{i\hat{L}(t-\tau)} i\hat{P}\hat{L}e^{i\hat{Q}\hat{L}\tau} \qquad (1.44)$$

It allows to represent the second term in equality (1.38) in the form:

$$e^{i\hat{L}t}F_m^Q = \int_0^t d\tau e^{i\hat{L}(t-\tau)} i\hat{P}\hat{L}e^{i\hat{Q}\hat{L}\tau}F_m^Q + e^{i\hat{Q}\hat{L}t}F_m^Q \ . \qquad (1.45)$$

The value

$$F_m^Q(t) \equiv e^{i\hat{Q}\hat{L}t} i\hat{Q}\hat{L}A_m \qquad (1.46)$$

by construction is always orthogonal to all values of $A_k$, i.e. $\langle A_k | F_m^Q(t) \rangle = 0$.

*Exercise.1.6.*

*Verify this statement by expanding the exponential operator into a Taylor series.*

By definition, $F_m^Q(t)$ is called the generalized stochastic force of Langevin associated with the quantity $A_m$.

Next, using formulas (1.29) and (1.46), the right side of equation (1.45) can be expressed as:

$$e^{i\hat{L}t}F_m^Q = \int_0^t d\tau \sum_{k,l} A_k(t-\tau)\left[\frac{1}{\langle A\otimes A\rangle}\right]_{k\,l} \left\langle A_l \middle| i\hat{L}F_m^Q(\tau)\right\rangle + F_m^Q(t) \qquad (1.47)$$

Let's consider the "ket" vector

$$\left|F_l^Q\right\rangle = i\hat{Q}\hat{L}\left|A_l\right\rangle \ . \qquad (1.48)$$

Hermitian adjoint to it is the "bra" vector

$$\left\langle F_l^Q\right| = -\left\langle A_l\right|\hat{L}\hat{Q}\,i \ . \qquad (1.49)$$

Furthermore, since $\hat{Q}$ is the projection operator onto the orthogonal complement of $\mathbf{L_n}$

$$\hat{Q}F_m^Q(\tau) = F_m^Q(\tau) \ . \qquad (1.50)$$

Equations (1.49) and (1.50) allow us to rewrite relation (1.47) as:

$$e^{i\hat{L}t}F_m^Q = -\int_0^t d\tau \sum_k K_{m\,k}(\tau)A_k(t-\tau) + F_m^Q(t), \qquad (1.51)$$

where

$$K_{m\,k}(t) \equiv \sum_l \left[\frac{1}{\langle A\otimes A\rangle}\right]_{k\,l} \left\langle F_l^Q \middle| F_m^Q(t)\right\rangle \ . \qquad (1.51a)$$

The values $K_{m\,k}(t)$ form what is called memory matrix $\hat{K}$:

$$\hat{K}(t) = \left\langle F^Q \otimes F^Q(t)\right\rangle \frac{1}{\langle A\otimes A\rangle} \qquad (1.52)$$

where $\left\langle F^Q \otimes F^Q(t)\right\rangle$ denotes a matrix whose elements are dynamic correlation functions of Langevin stochastic forces $\left\langle F_l^Q \middle| F_m^Q(t)\right\rangle$.

Thus, using ratios (1.39) and (1.51), equality (1.38) can be expressed as:

$$\frac{d}{dt}A_m(t) = \sum_k i\omega_{m\,k}A_k(t) - \int_0^t d\tau \sum_k K_{m\,k}(\tau)A_k(t-\tau) + F_m^Q(t) \qquad (1.53)$$

In fact, we have a system of n equations, since m=1,2,...,n. These equations can be given a compact matrix form, if the set $A_1(\Gamma), A_2(\Gamma), \ldots, A_n(\Gamma)$ is considered as an n-dimensional column vector A, and the set of corresponding generalized stochastic forces , $F_1^Q(t),...,F_n^Q(t)$, is considered as an n-dimensional column vector $F^Q(t)$ . Then the system of equations (1.53) can be written as a single matrix

$$\frac{d}{dt}A(t) = i\hat{\omega}A(t) - \int_0^t d\tau\hat{K}(\tau)A(t-\tau) + F^Q(t) \tag{1.54}$$

where

$$\hat{\omega} = \left\langle A \middle| \hat{L}A \right\rangle \frac{1}{\langle A \otimes A \rangle} \text{ -} \tag{1.54a}$$

frequency matrix,

$$\hat{K} = \left\langle F^Q(0)F^Q(\tau) \right\rangle \frac{1}{\langle A \otimes A \rangle} \text{ -} \tag{1.54b}$$

memory matrix

$$F^Q(\tau) \equiv e^{i\hat{Q}\hat{L}\tau}F^Q \text{ -} \tag{1.54c}$$

Langeven stochastic force.

The matrix equation (1.54) or the equivalent system (1.53) is called the generalized Langevin equation.

Thus, the generalized Langevin equation is a non-homogeneous (due to the presence of $F^Q(t)$ ) integro-differential relation for *A(t)*. The kernel of this equation, called the memory matrix, turns out to be proportional to the matrix of dynamic correlations of heterogeneous components—generalized Langevin stochastic forces. The matrix relation (1.54b), or the equivalent relation (1.54a) for matrix elements, between the memory matrix and the matrix of dynamic correlations of Langevin stochastic forces constitutes the content of the fluctuation-dissipation theorem.

The stochastic force of Langevin $F_k^Q(t)$ at any moment in time is orthogonal to $A_k(\Gamma)$, i.e. $\left\langle A_k \middle| F_k^Q(t) \right\rangle = 0$ . The evolution of stochastic force is determined not by real dynamics, i.e., the normal evolution operator, but by "projective" dynamics, i.e., the "anomalous" propagator (see (1.54c) or (1.46)).

As already noted, due to the presence of terms related to the stochastic Langevin force, the generalized Langevin equation is in fact a system of inhomogeneous integro-differential equations with respect to physical quantities "of interest." Without losing generality, it is possible to obtain homogeneous integro-differential equations from them in a simple way. The point is that for interpreting specific experimental results, it is not the instantaneous values of dynamic quantities that *Ak(t)* are important, but their dynamic equilibrium correlation functions

$$C_{m\,l}(t) = \langle A_l \,|\, A_m(t) \rangle \qquad (1.55)$$

forming a matrix of dynamic correlations $\hat{C}(t)$.

Multiplying the ratio (1.53) on the left by $\langle A_l |$ , due to the orthogonality of $F_m^Q(t)$ to $\langle A_l |$, we obtain the following homogeneous integro-differential equations:

$$\frac{d}{dt} C_{m\,l}(t) = \sum_k i\omega_{m\,k} C_{k\,l}(t) - \int_0^t d\tau \sum_k K_{m\,k}(\tau) C_{k\,l}(t-\tau) \qquad (1.56)$$

or, in matrix form:

$$\frac{d}{dt} \hat{C}(t) = i\hat{\omega}\hat{C}(t) - \int_0^t d\tau \hat{K}(\tau)\hat{C}(t-\tau) \ . \qquad (1.56b)$$

Note that any of the forms of the generalized Langevin equations (1.56a) or (1.56b) is an exact consequence of the microscopic equations of motion, i.e., Newton's equations or their equivalent Hamilton equations. However, for a truly mathematically correct formulation of the problem, it is necessary to know:

1) initial values of the dynamic correlation matrix at the initial moment in timeначальные $\hat{C}(0) = \left\| \langle A_k \,|\, A_l \rangle \right\|$.
2) memory matrix $\hat{K}(\tau) = \left\| \langle F_k^Q \,|\, F_l^Q(\tau) \rangle \right\|$.

Each of them reduces to a common unsolved problem of many-particle theory—the problem of closing an infinite system of coupled equations.

The calculation of the matrix $\hat{C}(0)$, as well as the calculation of the frequency matrix $\hat{\omega}$, can be reduced to the calculation of equilibrium correlation functions, which is the subject of equilibrium statistical mechanics. In problems of physical kinetics, this problem is considered to be already solved in the necessary approximation. Calculating the memory matrix $\hat{K}(\tau)$ is a key part of using the projection operator method to analyze specific physical problems.

First of all, it should be noted that the derivation of equations (1.56a) and (1.56b) is so general that they are equally applicable to systems with a small number of particles, whose behavior is reversible in time, and to systems with a macroscopically large number of particles, whose evolution is irreversible in time. This fundamental difference in behavior is reflected in the qualitatively different behavior of the memory matrix $\hat{K}(\tau)$ at $t \to \infty$. For a finite number of particles, there is no $\lim_{t\to\infty} \hat{K}(t)$, the memory matrix is a periodic or quasi-periodic function of time.

*Exercise.1.7.*

*Consider the model problem: let n=1 $\hat{K}(\tau)$=const. Verify that C(t) describes the behavior of a harmonic oscillator.*

For irreversible systems

$$\lim_{t\to\infty} \hat{K}(t) = 0 \tag{1.57}$$

i.e., the memory matrix should decay over time.

Ideally, the decay of the memory matrix should be proven as a consequence of the macroscopic nature, i.e., in the thermodynamic limit $N \to \infty$ ; $N/V = const$ , and special properties of the Hamiltonian $H(\Gamma)$ such as “ mixing,” leading to the dynamic “stochasticity” of the solutions to Hamilton’s equations. However, to now such program of justification of statistical physics has not been fully realized. For this reason, the irreversibility of the evolution of macroscopic systems, or the second law of thermodynamics, is actually postulated by means of some additional requirement such as “correlation weakening” or the hypothesis of “molecular chaos,” etc. . The relation (1.57) is in fact one form of such an additional requirement: memory loss is the source of irreversibility.

In addition to fulfilling this general requirement, it is necessary to have actual knowledge of the memory matrix as a function of time at any given moment t , otherwise equations (1.56a) and (1.56b) are not closed, since they contain new unknown quantities $K_{mk}(\tau)$the matrix elements of the matrix. As already mentioned, the problem of obtaining a closed system of a finite number of equations for a finite set of physical quantities is a general unsolved problem of statistical physics: equilibrium and non-equilibrium.

The most well-known way to formulate it is the infinite chain of BBGKY (Bogoliubov-Born-Green-Kirkwood-Yvon) for reduced equilibrium distribution functions. As is well known, the integral-differential equation for a binary distribution function $g_2(\vec{r}_{12})$ contains a three-partial function $g_3(\vec{r}_{12};\vec{r}_{23};\vec{r}_{13})$, the equation for which, in turn, contains a four-partial distribution function $g_4(\{\vec{r}_{ij}\})$ , and so on. Multi-particle correlations arising from interactions are the reason for this non-closedness.

A similar BBGKY chain arises for non-equilibrium reduced distribution functions. In our case of the generalized Langevin equation, the appearance of an unknown memory matrix $\hat{K}(\tau)$in relations (1.56a) and (1.56b) should be considered as a specific form of manifestation of the BBGKY chain. Thus, if we wanted to write dynamic equations for the memory matrix, we would have to supplement the quantities of interest to us with Langevin stochastic forces $f^Q(0) = \{f_1^Q(0), \quad f_2^Q(0), \quad \ldots, \quad f_N^Q(0)\}$. We could then define the projection operator $\hat{P}_1$ onto the corresponding linear subspace. Next, we can derive a new generalized Langevin equation for $f^Q(t)$ :

$$\frac{d}{dt} f^Q(t) = i\hat{\Omega}_1 f^Q(t) - \int_0^t d\tau \hat{K}_1(\tau) f^Q(t-\tau) + f_1^Q(t) \quad , \tag{1.58}$$

where

$\hat{\Omega}_1 = \left\langle f \otimes \hat{L}_1 f \right\rangle \dfrac{1}{\left\langle f^Q \otimes f^Q \right\rangle}$ - frequency matrix for first-order Langevin stochastic forces $f^Q(t)$,

$\hat{L}_1 = \hat{Q}\hat{L}$ - anomalous Liouville superoperator describing the dynamics аномальный $f^Q(t)$,

$f_1(t) \equiv \exp\{i(1-\hat{P}_1)\hat{L}_1 t\} i\hat{Q}_1 \hat{L}_1 \hat{f}$ - Langeven stochastic force of the “second” order,

$K_1(t) \equiv \langle f_1(0) \otimes f_1(t) \rangle \left\langle f^Q \otimes f^Q \right\rangle^{-1}$ - second-order memory matrix.

*Exercise 1.8.*

*1) Calculate explicitly the matrix elements of the frequency matrix $\hat{\Omega}_1$ and the memory matrix $\hat{K}_1(\tau)$.*

*2) Make sure that $\hat{P} \bullet \hat{P}_1 = 0$ . Recall that $\hat{P}$ is the projection operator onto the initial values $A_1(\Gamma)$, $A_2(\Gamma)$, ..., $A_n(\Gamma)$.*

The equation (1.58) obtained above allows, in principle, based on the fluctuation-dissipation theorem (see relations (1.54b), (1.51a)), to determine the memory matrix $\hat{K}(\tau)$. However, it itself contains a new unknown memory matrix $\hat{K}_1(\tau)$ of the “second” order. Thus, it becomes clear that mathematically, such a rigorous and consistent formulation of the generalized Langevin equation inevitably leads to an infinite chain of integro-differential equations. The problem of determining the memory matrix is essentially a problem of closing or opening the BBGKY chain for a hierarchical sequence of memory matrices.

The problem of closure can be considered relatively solved only for a relatively narrow range of tasks containing a clearly identified small parameter reflecting the small number of interactions. The most striking example is rarefied gases. The small parameter here is the value $n\sigma^3$, where $n$ is the concentration of molecules, $\sigma$ is the characteristic linear dimension, or "diameter," of the molecules. The potential energy of interactions is, on average, small here compared to kinetic energy. Other examples include crystal vibrations at temperatures much lower than the melting point. n fact, we are dealing with a dilute gas of phonons – quasi-particles that describe the collective vibrations of the crystal lattice. The energy of phonon interaction, corresponding to anharmonic contributions to the total Hamiltonian of the system, is also much smaller than the kinetic energy of phonons, i.e., the harmonic part of the Hamiltonian. The existence of a small parameter allows us to formulate a perturbation theory for this parameter, which in principle should lead to the possibility of controlled approximations, i.e., the problem turns out to be fundamentally solvable with any predetermined accuracy.

*Exercise 1.9.*

*Consider whether the small parameter in the case of a rarefied phonon gas can be represented as* $n\sigma^3$ *, which in this case will describe* $\sigma$ *. Note that if perturbation theory can be formulated, then the system is always nearly ideal.*

However, even in these simplest situations, it is relatively easy to consider only the lowest contributions to a small parameter, which is usually the second order of perturbation theory. Contributions of higher orders of perturbation theory often lead to divergent integrals when interpreted straightforwardly, and an additional problem of regularization arises, which is solved in each specific case taking into account the specific features of the problem.

In cases where interactions are not small, the definition of the memory matrix involves the introduction of certain additional postulates, which, at best, are based only on some "physical" considerations or heuristic ideas. From a logically consistent point of view, these additional postulates are uncontrollable approximations in relation to the exact dynamic equations of motion.

The final criterion for the adequacy of the approximations adopted in this situation is a comparison of the predictions of the "theory" with experimental data. Essentially, this area of theoretical physics is the art of finding the best approximation.

In our opinion, the best illustration of the above is a reminder of the real state of the problem of closing the BBGKY chain for the simplest problem: determining the equilibrium binary radial distribution function $g_2(\vec{r})$. The three following approximations are the most well-known here:

Superposition approximation (SA), linking a three-part distribution function $g_3(\vec{r}_{12};\vec{r}_{23};\vec{r}_{13})$ with a binary one:

$$g_3(\vec{r}_{12};\vec{r}_{23};\vec{r}_{13}) = g_2(\vec{r}_{12})\, g_2(\vec{r}_{13})\, g_2(\vec{r}_{23}). \tag{1.59a}$$

2. Hyperchain approximation (HC), linking a direct correlation function $C(\vec{r})$ with $g_2(\vec{r})$ and the intermolecular interaction potential $U(\vec{r})$ :

$$C(\vec{r}) = g_2(\vec{r}) - 1 - \ln g_2(\vec{r}) - \frac{U(\vec{r})}{kT}. \tag{1.59b}$$

3. Percus-Yevick approximation (PY):

$$C(\vec{r}) = g_2(\vec{r}) \exp\left\{\frac{U(\vec{r})}{kT}\right\}\left(\exp\left\{-\frac{U(\vec{r})}{kT}\right\} - 1\right). \tag{1.59c}$$

From a purely theoretical point of view, all three approximations are equivalent, since none of them is derived as the result of some systematic expansion in a small parameter. They are, in fact, guessed on the basis of "truthful" arguments.

The most physically transparent arguments are those underlying the SP approximation: relation (1.59a) is equivalent to approximating the effective potential of three-particle interaction through effective pair interaction potentials. The physical commentary on relations (1.59b) and (1.59c), HC and PY, is not so obvious and cannot be made in one or two sentences (interested readers are referred to the relevant literature).

Although it has the greatest physical transparency, the SP approximation is the worst when compared with the equation of state of a liquid of hard spheres obtained on the basis of numerical experiments. In this respect, the PI approximation is the best. Meanwhile, the superposition approximation is the only one of those discussed that predicts a liquid-crystal phase transition.

When it comes to defining the memory matrix of the generalized Langevin equation, it is important to understand that the situation only becomes more complicated:

1) At the initial moment of time t=0, it is necessary to be able to calculate the equilibrium many-body correlation functions.

2) The temporal evolution is determined by the "projection" dynamics, $\exp\left\{i\hat{Q}\hat{L}Qt\right\}$ .

In this point, the reader should clearly understand that they are dealing with an unresolved general scientific problem and can contribute to its resolution themselves.

### 1.2.4. Generalisation to quantum systems.

The formalism proposed above can be easily extended to quantum systems. Physical quantities in quantum mechanical descriptions correspond to certain self-adjoint operators acting in the Hilbert space of wave functions.

The set of all linear operators is itself a linear space over the field of complex numbers. Between two quantum mechanical operators, a scalar product can be defined in the sense of Kubo:

$$\left\langle \hat{A} \middle| \hat{B} \right\rangle \equiv \beta^{-1} \int_0^\beta d\lambda Spur\left( \hat{A}^* \hat{B}(i\hbar\lambda) \hat{\rho}_{eq} \right), \tag{1.60}$$

where $\hat{A}^*$ - operator, Hermitian conjugate to $\hat{A}$,

$\hat{B}(\tau) \equiv \exp\left\{ i \dfrac{\hat{H}\tau}{\hbar} \right\} \hat{B} \exp\left\{ -i \dfrac{\hat{H}\tau}{\hbar} \right\}$ - operator $\hat{B}$ in Heisenberg's representation a,

$\beta = \dfrac{1}{kT}$ - inverse temperature , $\hat{\rho}_{eq}$ - the equilibrium density matrix of all system.

After introducing the metric (1.60), the set of all linear operators forms a Lieuwille space. The Heisenberg equation for the operator $\hat{A}(t)$ has the

$$\frac{d}{dt} \hat{A}(t) = \frac{i}{\hbar} \left[ \hat{H}; \hat{A}(t) \right] \equiv i\hat{L}\hat{A}, \tag{1.60a}$$

where $\hat{L} = \dfrac{1}{\hbar} \left[ \hat{H}; \ldots \right]$ is the Liouville superoperator.

We provide the reader with the opportunity to show that all of the above calculations can be repeated and that the generalized Langevin equation (1.56a) and (1.56b) has a common structure for quantum and classical systems.

### 1.3. Some applications.

### 1.3.1. Brownian motion.

Let us consider a “test” particle of mass M placed in a medium consisting of N classical particles of mass m. The total Hamiltonian of the entire system is as follows:

$$H(\Gamma)=\frac{\vec{P}^2}{2M}+\sum_i V(\vec{R}-\vec{r}_i)+\sum_i \frac{\vec{p}_i^2}{2m}+\sum_{i<j} U(\vec{r}_{ij}), \tag{1.61}$$

where

$\vec{P}$ and $\vec{r}_i$ – momentum and radius vector of the test particle,

$\vec{p}_i$ and $\vec{r}_i$ – momentum and radius vector of the i-th particle of the medium,

$V(\vec{R}-\vec{r}_i)$ – potential energy of interaction between the test particle and the i-th particle of the medium,

$U(\vec{r}_{ij})$ – potential energy of interaction between two particles of the medium.

When the condition $M >> m$ is met, the test particle is referred to as a Brownian particle, in memory of Robert Brown, an English botanist who first observed the random, continuous motion of pollen particles suspended in liquid in 1829.

Let us write down the generalized Langevin equation for the momentum of the test particle, i.e., as quantities "of interest," we will choose the components of the momentum of the test particle: $A_1 = p_x$, $A_2 = p_y$ , $A_3 = p_z$.. The components of the momentum are orthogonal vectors in the Liouville space **L**, and the projection operator onto the subspace spanned by them is equal to:

$$\hat{P}=|\vec{p}\rangle\frac{1}{MkT}\langle\vec{p}| \tag{1.62}$$

Using the definition of the frequency matrix (see relation (1.39a)), it is easy to verify that in this case $\hat{\omega}=0$ . Indeed, the Hamiltonian (1.61) is invariant under time reversal. The Leeville Superoperator

$$\hat{L} = i\{H,...\} = i\sum_i \left( \frac{\partial}{\partial \vec{r}_i} H \frac{\partial}{\partial \vec{p}_i} - \frac{\partial}{\partial \vec{p}_i} H \frac{\partial}{\partial \vec{r}_i} \right) +$$

$$+ i\frac{\partial}{\partial \vec{R}} H \frac{\partial}{\partial \vec{P}} - i\frac{\partial}{\partial \vec{P}} H \frac{\partial}{\partial \vec{R}} = \quad , \qquad (1.63)$$

$$= \frac{1}{i}\left[ \sum_k \left( \frac{1}{m}\vec{p}_k \frac{\partial}{\partial \vec{r}_k} + \vec{f}_k \frac{\partial}{\partial \vec{p}_k} \right) + \frac{1}{M}\vec{P}\frac{\partial}{\partial \vec{R}} + \vec{F}\frac{\partial}{\partial \vec{P}} \right]$$

where $\vec{f}_k$- total force acting on the k-th particle of the, $\vec{F}$- is the total force acting on the test particle . The Leeville Superoperator is odd in time reversal operations.

нечётен при операции обращения во времени. Therefore, the matrix

$$\left\langle \vec{p} \otimes \hat{L}\vec{p} \right\rangle = \left\| \left\langle p_\alpha \hat{L} p_\beta \right\rangle_{eq} \right\| = 0$$

and (see формулы (1.39a) and (1.40) brings equality $\hat{\omega} = 0$.

The generalized stochastic force of Langevin at the initial moment of time can be calculated based on formula (1.41):

$$\vec{F}^Q = i\hat{Q}\hat{L}\vec{p} = i(1-\hat{P})\hat{L}\vec{p} = \vec{F} = -\sum_i \frac{\partial}{\partial \vec{R}} V(\vec{R} - \vec{r}_i) \qquad (1.64)$$

We see that at the initial moment of time it coincides with the total force with which the particles of the medium act on the test particle. At any given moment of time, the stochastic Langevin force acting on the particle is determined by the ratio (see formula (1.46)):

$$\vec{F}^Q(t) = \exp\left\{ i\hat{Q}\hat{L}t \right\} \vec{F} \qquad (1.64a)$$

The memory matrix, defined by the general ratio (1.52), is diagonal in this case due to the assumed isotropy of the system:

$$K_{\alpha\beta}(t) = \frac{1}{MkT}\left\langle F_\alpha^Q(0) F_\alpha^Q(t) \right\rangle \delta_{\alpha\beta} = \frac{1}{3MkT}\left\langle \vec{F}(0)\vec{F}^Q(t) \right\rangle \delta_{\alpha\beta} \,. \qquad (1.65)$$

Thus, the generalized Langevin equation for the case under discussion takes the form:

$$\frac{d}{dt}\vec{p}(t) = -\frac{1}{3MkT}\int_0^t \left\langle \vec{F}^Q(0) \middle| \vec{F}^Q(\tau) \right\rangle \vec{p}(t-\tau)\,d\tau + \vec{F}^Q(t) \ . \qquad (1.66)$$

*Exercise 1.10.*

*As quantities of interest, we could choose* $\{A_1, A_2, A_3, A_4, A_5, A_6\} = \{\vec{P}, \vec{R}\}$ *. Write down the generalized Langevin equation for this case. Make sure that the frequency matrix in this case is not zero; additional relations are added to (1.66):*

$$\frac{d}{dt}\vec{R}(t) = \frac{1}{M}\vec{P}(t) \ . \qquad (1.66a)$$

For a Brownian particle, when M>>m, the integral-differential equations (1.66) are simplified based on the following qualitative physical arguments. Typical values of the momentum of a Brownian particle $\left\langle \left|\vec{P}\right| \right\rangle \cong \sqrt{MkT}$, typical values of the momenta of the medium particles $\left\langle \left|\vec{p}_i\right| \right\rangle \cong \sqrt{mkT}$, $\sqrt{MkT} >> \sqrt{mkT}$. The particles of the medium are much more mobile than Brownian particles, since their velocities are of the order of $\sqrt{\frac{kT}{m}} >> \sqrt{\frac{kT}{M}}$. Therefore, it is natural to assume that at typical times of attenuation of dynamic correlations of Langevin's stochastic force $\left\langle \vec{F}(0)\vec{F}^Q(t) \right\rangle$ the momentum of a Brownian particle changes little. This allows us to apply the Markov approximation to the integral-differential equation (1.66), i.e., replace $\vec{p}(t-\tau)$ with $\vec{p}(t)$ and set the upper limit of integration equal to infinity. As a result, we obtain a differential stochastic equation:

$$\frac{d}{dt}\vec{p}(t) = -\zeta\,\vec{V}(t) + \vec{F}^Q(t) \qquad (1.67)$$

where

$$\zeta = \frac{1}{3kT}\int_0^\infty \left\langle \vec{F}^Q(0)\vec{F}^Q(t)\right\rangle dt \text{ -} \tag{1.67a}$$

is the friction coefficient of a Brownian particle relative to its environment.

Equation (1.67) is called the Langevin equation. It was proposed on the basis of phenomenological considerations by P. Langevin in 1908 to explain Brownian motion. The Zwanzig-Mori projection operator technique, which appeared in the early 1960s, allows us to give the most consistent microscopic "derivation" of Langevin's stochastic equations at the current level of development of modern theory of irreversible processes.

Let us discuss the main stages of this derivation once again. The integral-differential equation, which is actually a relation, is an exact consequence of Hamilton's microscopic equations. However, it is not closed, since it contains a new unknown quantity- memory, which is a dynamic correlation function of a non-homogeneous term $\vec{F}^Q(t)$, called, by definition, the Langevin stochastic force. Furthermore, based on physical considerations, this equation is then closed by applying (in fact, postulating) the Markov approximation for the unknown memory function:

$$\frac{1}{3MkT}\left\langle \vec{F}^Q(t_1)\vec{F}^Q(t_2)\right\rangle = 2\zeta\,\delta(t_2 - t_1) \tag{1.68}$$

The friction coefficient $\zeta$ is obtained in accordance with the relation (1.67a) as the integral of the memory function.

The quantitative basis for approximation (1.68) is the presence of a small parameter $\sqrt{\frac{m}{M}}$, or, equivalently, the presence of rapid movements of medium particles compared to Brownian particles. The Langevin stochastic force is interpreted as a normal delta-correlated random process.

### 1.3.2. Markov approximation: general case.

The situation described above can be easily generalized if the physical quantities of interest to us, $A_1(\Gamma)$, $A_2(\Gamma)$, ..., $A_n(\Gamma)$, are “slower” than all the others that determine the attenuation of the memory matrix $\hat{K}(t)$. In other words, the characteristic time of correlations of dynamic correlation functions determining the dynamics of the memory matrix is much shorter than the characteristic relaxation time of quantities. In this case, as with the “derivation” of the Langevin equation, the Markov approximation can be applied:

$$K_{mk}(t) = 2w_{mk}\delta(t) \tag{1.69a}$$

where the coefficients

$$w_{mk} = \int_0^\infty K_{mk}(t)dt \tag{1.69b}$$

by definition are called kinetic coefficients and form a matrix of kinetic coefficients, or relaxation matrix.

The generalized Langevin equations (1.53) or (1.56) from integro-differential equations are simplified and become differential kinetic equations:

$$\frac{d}{dt}A_m(t) = \sum_k \left(i\omega_{mk} - w_{mk}\right)A_k(t) + F_m^Q(t) \tag{1.70a}$$

or

$$\frac{d}{dt}C_{ml}(t) = \sum_k \left(i\omega_{mk} - w_{mk}\right)C_{kl}(t)\ . \tag{1.70b}$$

The matrix of kinetic coefficients $\hat{w} = \left\| w_{km} \right\|$, often called the relaxation matrix, turns out to be the integral of the memory matrix:

$$\hat{w} = \int_0^\infty \hat{K}(\tau)d\tau \tag{1.71}$$

Relationships (1.71) are called Green-Kubo relationships for kinetic coefficients. Relationships similar in structure can be obtained using the Kubo-Tomita method of linear response to external perturbations. However, there is also a difference: in the Kubo-Tomita method, the dynamic correlation functions contained in the subintegral expressions of relation (1.71) evolve under the action of the real propagator, i.e. $\exp\left\{i\hat{L}t\right\}$, the superoperator , while in relation (1.71), evolution occurs, as already noted, under the action of the "projection" superoperator of evolution $\exp\left\{i\hat{Q}\hat{L}Qt\right\}$ .

The approximation

$$\exp\left\{i\hat{Q}\hat{L}Qt\right\} \cong \exp\left\{i\hat{L}t\right\} \tag{1.72}$$

is called the self-consistent approximation.

In a number of problems, such as deriving heat conduction equations, hydrodynamic equations, self-diffusion, etc., it is rigorously proven that approximation (1.72) is asymptotically accurate in the limit of small wave vectors or when $t \to \infty$ . In general, the problem of the relationship between real dynamics and projection dynamics is one of the pressing unsolved problems in the theory of irreversible processes.

In the case where the entire system is a system of harmonic oscillators, the projection dynamics turns out to be Hamiltonian with a Hamiltonian in which the kinetic energy of the subsystem for which the kinetic equations are written is removed from the total Hamiltonian of the system. This can already be seen from relation (1.66). Indeed, the initial value of the stochastic Liegevin force $\vec{F}^{Q}(t=0)$ in this case is a certain linear function of the differences between the coordinates of the particles. For this reason, it turns out that $\exp\left\{i\hat{Q}\hat{L}Qt\right\}\vec{F}^{Q}(0) = \exp\left\{i\hat{L}_{\mu}t\right\}\vec{F}^{Q}(0)$ , where $\hat{L}_{\mu}$ is the Liouville superoperator of the matrix. This issue will be discussed in more detail in the second chapter.

Let us now formulate the general conditions for the applicability of the Markov approximation, sometimes referred to as the short-time correlation approximation. Consider the matrix of dynamic correlations of Langevin stochastic forces:

$$\hat{D}(t) = \left\| \left\langle F_m^Q(0) \middle| F_k^Q(t) \right\rangle \right\| \tag{1.73}$$

Let's make it “dimensionless.” To do this, let's consider its value at t=0, $\hat{D}(0) \equiv \hat{D}$ which is non-degenerate in all physically interesting cases, i.e., exists $\hat{D}^{-1}$. Matrix

$$\hat{\tilde{D}}(t) = \hat{D}^{-1}\hat{D}(t) \tag{1.74}$$

s already dimensionless. Consider the integral over time $\hat{\tilde{D}}(t)$:

$$\hat{\tau} \equiv \int_0^\infty \hat{\tilde{D}}(t)dt \tag{1.75}$$

The matrix elements of the matrix $\hat{\tau}$ can be considered as characteristic correlation times of our system. The matrix of kinetic coefficients $\hat{w}$ determines the rate of relaxation processes, which must be small compared to the decay rate of the memory matrix. It is easy to see that this is equivalent to the requirement

$$\hat{w}\,\hat{\tau} << 1 \tag{1.76}$$

that is, all matrix elements of the product are small compared to unity.

*Exercise (non-trivial)1.11.*

*Based on the general equations (1.70a) and (1.70b), derive Bloch's equations describing the relaxation of the components of the paramagnetic magnetic moment in an external magnetic field.*

### 1.3.3. A simple example of non-Markovian behavior.

Let us return to equation (1.66), which describes the motion of a test particle in a medium. If the mass of this particle is small, i.e., it cannot be considered large compared to the mass of the medium particles, then we have no reason to use the Markov approximation (1.68). Let us consider a model non-

Markovian situation that yields non-trivial results: the case of exponential decay of the memory matrix:

$$K(\tau) = K_0 \exp\left\{-\frac{\tau}{\tau_0}\right\} \qquad (1.77)$$

where $\tau_0$ - memory matrix decay time,

$K_0 = \frac{1}{3kTM}\left\langle \vec{F}^2(0) \right\rangle$ - initial memory value.

Multiply both sides of equation (1.66) by the initial value of the particle's velocity $\vec{V} = \frac{1}{M}\vec{P}$, and divide by the mass M. As a result, we obtain the following equation for the velocity-velocity autocorrelation $C_V(t) \equiv \left\langle \vec{V}(t)\vec{V}(0) \right\rangle$:

$$\frac{d}{dt}C_V(t) = -\int_0^t K_0 \exp\left\{-\frac{\tau}{\tau_0}\right\} C_V(t-\tau)d\tau \quad . \qquad (1.78)$$

Equation (1.78) can be easily solved using Laplace transforms: the right-hand side of (1.78) is the convolution of two functions $\mathrm{K}(\tau) \equiv \mathrm{K}_0 \exp\left\{-\frac{\tau}{\tau_0}\right\}$ and $C_V(t-\tau)$. Therefore, Laplace transforms equation (1.78) has a simple form:

$$\mathrm{p}\hat{\mathrm{C}}_\mathrm{V}(\mathrm{p}) - \mathrm{C}_\mathrm{V}(0) = -\hat{\hat{\mathrm{K}}}(\mathrm{p})\hat{\mathrm{C}}_\mathrm{V}(\mathrm{p}) \ , \qquad (1.79)$$

where $\hat{f}(p) = \int_0^\infty f(t)\exp\{-pt\}dt$ – standard notation for Laplace transform of a function $f(t)$, $C_V(0) = \left\langle V^2 \right\rangle$ – mean square value of the velocity of a test particle.

Equation (1.79) is a linear equation for $\hat{C}_V(p)$. Solving it, we find:

$$\hat{\mathrm{C}}_\mathrm{V}(\mathrm{p}) = \frac{\left\langle \mathrm{V}^2 \right\rangle}{\mathrm{p} + \hat{\hat{\mathrm{K}}}(\mathrm{p})} \qquad (1.80)$$

We find the correlation function $C_V(t)$ we are interested in using Mellin's formula:

$$C_V(t) = \frac{1}{2\pi i}\int_{x-i\infty}^{x+i\infty} \exp\{pt\}\hat{C}_V(p)dp \quad . \qquad (1.81)$$

The Laplace transform of a memory function is easily calculated:

$$\hat{\tilde{K}}(p) = \frac{\tilde{K}_0}{p+\frac{1}{\tau_0}} \qquad (1.82)$$

Next, using relations (1.80), (1.81) and Cauchy's theorem on residuals, we find the correlation function we are interested

$$C_V(t) = \langle V^2\rangle \exp\left\{-\frac{t}{2\tau_0}\right\}\left\{ch\left(\frac{\sqrt{1-4\tilde{K}_0\tau_0^2}}{2\tau_0}t\right) + \frac{sh\left(\frac{\sqrt{1-4\tilde{K}_0\tau_0^2}}{2\tau_0}t\right)}{\sqrt{1-4\tilde{K}_0\tau_0^2}}\right\} \qquad (1.83)$$

*Exercis1.12. Restore the omitted calculations between ratios (1.80) and (1.83).*

We see that the decay of the autocorrelation function $C_V(t)$ is generally non-exponential. Moreover, the behavior changes qualitatively depending on the value of the dimensionless parameter $4\tilde{K}_0\tau_0^2$ . Three regimes can be distinguished.

1.Short correlation times, when

$$4\tilde{K}_0\tau_0^2 << 1 \qquad (1.84)$$

Expanding the hyperbolic sine and cosine on the right side of equation (1.83), we obtain:

$$C_V(t) = \langle V^2\rangle\left[\exp\{-\tilde{K}_0\tau_0 t\}(1+\tilde{K}_0\tau_0^2) - \exp\left\{-\frac{t}{\tau_0}\right\}\tilde{K}_0\tau_0^2\right] \qquad (1.85)$$

At times $t<<\tau_0$, the decay is non-exponential, but overall $C_V(t)$, due to relation (1.84), it changes little. At $t >> \tau_0$, the first exponent dominates in relation (1.85), decaying with a characteristic relaxation time.

$$\tau_V = \frac{1}{\tilde{K}_0\tau_0} = \frac{3kTm}{\tau_0\left\langle \vec{F}^2\right\rangle} \tag{1.86}$$

It is easy to see that this limit of short correlation times, accurate to values of the order of , coincides with the results of the Markov approximation discussed in Section 3.1.

*Exercise 1.13.*

*Based on the results of sections 3.1. and 3.3., find a "microscopic" expression for the friction coefficient. What motion corresponds to the behavior of* $C_V(t)$ *at* $t << \tau_0$ *? What physical interpretation can you give to time* $\tau_0$ *?.*

2. When $\tilde{K}_0\tau_0^2 = 1/4$, the decay of the correlation function is non-exponential, and the characteristic decay time of $C_V(t)$ is twice as long as the decay time of the autocorrelation function of the Langevin force $\tau_0$. However, if relaxation time is defined as the integral over time of $C_V(t)/C_V(0)$

$$\tau_V \equiv \int_0^\infty \frac{C_V(t)}{C_V(0)}dt = \int_0^\infty \exp\left\{-\frac{t}{2\tau_0}\right\}\left\{1+\frac{t}{2\tau_0}\right\}dt = 4\tau_0 . \tag{1.87}$$

3. Strongly non-Markovian limit

$$4\tilde{K}_0\tau_0^2 >> 1 \tag{1.88}$$

In that limit, the under-radical expressions in relation (1.83) become negative, the hyperbolic sine and cosine become ordinary, and the behavior $C_V(t)$ can be characterized as damped oscillations:

$$C_V(t) = \left\langle V^2\right\rangle \exp\left\{-\frac{t}{2\tau_0}\right\}\left\{\cos\omega_0 t + \frac{\sin\omega_0 t}{2\omega_0\tau_0}\right\} , \tag{1.89}$$

where $\omega_0 = \sqrt{\tilde{K}_0} = \sqrt{\frac{\langle \vec{F}^2(0) \rangle}{3kTm}}$.

The characteristic frequency $\omega_0$ can be considered as the frequency of local oscillations. Note that the characteristic decay time 2 is longer than the decay time of the memory function. The formula (1.88) qualitatively correctly describes the attenuation of the autocorrelation function $C_V(t)$ in real liquids, which are characterized by the presence of damped oscillations. If the characteristic relaxation time is determined by analogy with (1.86a), then

$$\tau_V \equiv \int_0^\infty \frac{C_V(t)}{C_V(0)} dt = \int_0^\infty \exp\left\{-\frac{t}{2\tau_0}\right\}\left\{\cos\omega_0 t + \frac{\sin\omega_0 t}{2\omega_0\tau_0}\right\} \\ = \frac{4\tau_0}{1+4\omega_0^2\tau_0^2} \approx \frac{1}{\omega_0^2\tau_0}. \quad (1.90)$$

*Exercise 1.14.*

1. *Prove the exact ratio*

$$\langle r^2(t) \rangle = 2\int_0^t (t-\tau)\langle \vec{V}(\tau)\vec{V}(0) \rangle d\tau \quad (1.91)$$

*where* $\langle r^2(t) \rangle$ *- mean squared displacement of particle.*

2. *Using the results obtained above, verify that the diffusion behavior*

$$\langle r^2(t) \rangle = 6Dt \quad (1.92)$$

*has a place in any case at* $t \to \infty$ *, even in a strongly non-Markovian limit (1.89). Note that this diffusion regime occurs only when the autocorrelation function* $C_V(\tau) = \langle \vec{V}(\tau)\vec{V}(0) \rangle$ *decays sufficiently rapidly.*

3. *Derive the Kubo-Green formula*

$$D = \frac{1}{3}\int_0^\infty \langle \vec{V}(t)\vec{V}(0) \rangle dt \quad (1.93)$$

4. *Ensure that Einstein's ratio*

$$D = \frac{kT}{\zeta} \quad , \qquad (1.94)$$

*where* $\zeta = \frac{1}{3kT}\int_0^\infty \left\langle \vec{F}^Q(0)\vec{F}^Q(t) \right\rangle dt$

*is true in all cases, not just for Brownian particles.*

# Chapter 2.

## 2.1 Phase density of the probe macromolecule.

Let us assume that we are dealing with a polymer system and are interested in the dynamics of a certain macromolecule that has been isolated in some way. We will refer to this macromolecule as the "test" or "probe" macromolecule. We will refer to the rest of the system as the "matrix." The 'test' macromolecule moves in the "matrix," which is the "environment" relative to the selected chain.

Let denote the radius vectors and momenta $\vec{r}_1, \vec{r}_2, \ldots \vec{r}_N, \vec{p}_1, \vec{p}_2, \ldots \vec{p}_N$ of the segments of the probe macromolecule, and N denote the number of segments in it. Let's introduce a notation

$$\gamma_N \equiv \{\vec{r}_1, \vec{r}_2, \ldots \vec{r}_N, \vec{p}_1, \vec{p}_2, \ldots \vec{p}_N\} \ . \qquad (2.1)$$

The set $\gamma_N$ specifies a point in the phase space of the test macromolecule $P_N$ . The phase space of the entire system **P** is the tensor product of the phase spaces of all particles in the system. The points in space P, defined by the radius vectors and momenta of all particles, will be denoted by $\gamma$. Space $P_N$ is one of the factors of **P**.

We define the phase density of the probe macromolecule by the ratio:

$$f(\Gamma_N;\gamma_N) \equiv \delta(\Gamma_N - \gamma_N) \equiv \prod_{i=1}^{N} \delta\left(\vec{R}_i - \vec{r}_i\right)\delta\left(\vec{P}_i - \vec{p}_i\right), \qquad (2.2)$$

where $\Gamma_N \equiv \left\{\vec{R}_1, \vec{R}_2, \ldots, \vec{R}_N; \vec{P}_1, \vec{P}_2 \ldots, \vec{P}_N\right\}$, $\delta$ – Dirac delta function.

The arguments of phase density are the variables that form $\gamma_N$ . The variables $\Gamma_N$ that make up form a multidimensional index and are called field variables.. It is useful to draw an analogy with the " of interest" quantities $A_k(\Gamma)$ discussed in Section 1.2.3. The functions $A_k(\Gamma)$ are defined on phase space Γ, Γ is an argument, k=1,…,n is a discrete index. In the case of phase densities $f(\Gamma_N;\gamma_N)\equiv f_{\Gamma_N}(\gamma_N)$, $\Gamma_N$ plays a role of k, and $\gamma$ , in fact only $\gamma_N$ , is an argument. For brevity, we will use the following notation in the future:

$$f(\Gamma_N;\gamma_N)\equiv f(\Gamma_N), \tag{2.3}$$

omitting the indication of arguments, as in the case of $A_k$.

Average value $f(\Gamma_N)$

$$\langle f(\Gamma_N)\rangle \equiv \langle f(\Gamma_N)|1\rangle = \rho_N^*(\Gamma_N) \tag{2.4}$$

is a single-chain distribution function according to the coordinates and momenta of the probe macromolecule.

We define the effective single-chain Hamiltonian, or the effective free energy of the test macromolecule, by the relation:

$$H_N^* \equiv H_N^*(\Gamma_N) = -kT\ln\rho_N^*(\Gamma_N). \tag{2.5}$$

The effective Hamiltonian can be represented as:

$$H_N^* = \sum_{i=1}^{N}\frac{P_i^2}{2m} + W^*\left(\left\{\vec{R}_i\right\}\right), \tag{2.6}$$

where $W^*\left(\left\{\vec{R}_i\right\}\right)$ by definition is called effective intramolecular potential energy, or mean force potential.

The effective Hamiltonian $H_N^*$ differs from the exact single-chain Hamiltonian in that it contains effective or "renormalized" intramolecular interactions instead of true, "bare" intramolecular interactions. In fact, $W^*\left(\left\{\vec{R}_i\right\}\right)$ it is a complex multiparticle object that takes into account all multiparticle static correlations in an averaged form. The calculation $W^*\left(\left\{\vec{R}_i\right\}\right)$ is a non-trivial problem in statistical physics. Essentially (see relations (2.5) and (2.6)), knowledge of $W^*\left(\left\{\vec{R}_i\right\}\right)$ uniquely determines the single-chain configuration distribution of a test macromolecule in a given environment.

Consider an arbitrary physical quantity $A(\gamma)$ defined throughout the phase space of the system. Calculate the scalar product

$$\left\langle f(\Gamma_N)\middle|A(\gamma)\right\rangle = \int d\gamma\,\delta(\Gamma_N-\gamma_N)\rho_{eq}(\gamma)A(\gamma) = A^*(\Gamma_N)\rho_N(\Gamma_N) \qquad (2.7)$$

where

$$A^*(\Gamma_N) \equiv \frac{1}{\rho_N^*(\Gamma_N)}\int d\gamma\,\delta(\Gamma_N-\gamma_N)\rho_{eq}(\gamma)A(\gamma) \qquad (2.7a)$$

the average value of a physical quantity $A(\gamma)$ at fixed values of the coordinates and momenta of the test macromolecule, determined by the set $\Gamma_N$.

In particular, if $A(\gamma)$ depends only on the variables of the test macromolecule, then from relations (2.7) and (2.7a) it follows that

$$A^*(\Gamma_N) = A(\Gamma_N) \ . \qquad (2.8)$$

Let us calculate the scalar product of two densities $f(\Gamma_N)$ and $f(\Gamma_N')$ (see definition (1.1)):

$$\left\langle f(\Gamma_N)\middle|f(\Gamma_N')\right\rangle = \int d\gamma\rho_{eq}(\gamma)\delta(\Gamma_N-\gamma_N)\delta(\Gamma_N'-\gamma_N) =$$

$$= \rho_N^*(\Gamma_N)\delta(\Gamma_N-\Gamma_N') \ . \qquad (2.9)$$

This relation shows that the phase densities of the probe macromolecule are orthogonal to each other. Furthermore, the values $\left\langle f(\Gamma_N)\middle|f(\Gamma_N')\right\rangle$ can formally be regarded as matrix elements of an infinite matrix with indices $\Gamma_N$ and $\Gamma_N'$ . Equation (2.9) then shows that the matrix $\left\langle f(\Gamma_N)\middle|f(\Gamma_N')\right\rangle$ is proportional to the identity matrix defined by the multidimensional $\delta$ Dirac function $\delta\left(\Gamma_N - \Gamma'_N\right)$ . It is easy to calculate the inverse matrix of $\left\langle f\left(\Gamma_N\right) \middle| f\left(\Gamma'_N\right)\right\rangle$:

$$\left\langle f\left(\Gamma_N\right) \middle| f\left(\Gamma'_N\right)\right\rangle^{-1} = \frac{1}{\rho_N^*(\Gamma_N)}\delta\left(\Gamma_N - \Gamma'_N\right). \tag{2.10}$$

Consider the set of all phase densities $f(\Gamma_N)$ . The linear space spanned by all $f(\Gamma_N)$ is a subspace of the Liouville space of the entire system L. Let us consider the closure of this subspace in the topology of the Liouville space. Let us denote it by $L_N$ . It is easy to verify that L coincides with the Liouville space defined on the phase space of the test macromolecule. Let us consider the closure of this subspace in the topology of the Liouville space. Let us denote it by $L_N$. It is easy to verify that $L_N$ coincides with the Liouville space defined on the phase space of the probe macromolecule.

The projection operator on $L_N$ (cf. general definition (1.29)) is given by the relation:

$$\hat{P}_N = \int \left| f\left(\Gamma_N\right)\right\rangle \frac{d\Gamma_N d\Gamma'_N}{\left\langle f\left(\Gamma_N\right) \middle| f\left(\Gamma'_N\right)\right\rangle} \left\langle f\left(\Gamma'_N\right)\right| \quad . \tag{2.11}$$

By substituting ratio (2.10) into (2.11), we obtain:

$$\hat{P}_N = \int \left| f\left(\Gamma_N\right)\right\rangle \frac{d\Gamma_N}{\rho_N^*\left(\Gamma_N\right)} \left\langle f\left(\Gamma_N\right)\right|. \tag{2.12}$$

The ket vector $\left| f\left(\Gamma_N\right)\right\rangle$ in this formula, according to definition (2.2), is a Dirac $\delta$-function, which allows us to give it a compact form

$$\hat{P}_N = \frac{1}{\rho_N^*\left(\gamma_N\right)} \left\langle f(\Gamma_N = \gamma_N)\right| \quad . \tag{2.13}$$

Let a certain quantity $A(\gamma_N)$ depend only on the variables of the test macromolecule, i.e. $A(\gamma_N) \in L_N$ . Using relations (2.8) and (2.13), it is easy to verify that

$$\hat{P}_N A(\gamma_N) = A(\gamma_N), \tag{2.14}$$

i.e., indeed, the operator $\hat{P}_N$ is a projection operator on $L_N$.

*Exercise 2.1.*

*Calculate the projection of the total Hamiltonian of the entire system H(γ) onto $L_N$:*

$$\hat{P}_N H(\gamma) = -kT^2 \frac{\partial}{\partial T} \ell n \rho^*(\gamma_N) = H_N^*(\gamma_N) - T\frac{\partial}{\partial T} W^*(\gamma_N). \tag{2.15}$$

## 2.2. Generalized Langevin equation for phase densities.

Phase densities $f(\Gamma_N) = \delta(\Gamma_N - \gamma_N)$ are vectors in the Liouville space of the entire system **L**. Acting on them by the superoperator of evolution, we obtain phase densities in Heisenberg representation:

$$f(\Gamma_N;t) \equiv \exp\{i\hat{L}_\gamma t\} f(\Gamma_N) = \exp\{i\hat{L}_\gamma t\}\delta(\Gamma_N - \gamma_N) = \delta(\Gamma_N - \gamma_N(t)), \tag{2.16}$$

гд where $\hat{L}_\gamma = i\{H(\gamma);...\}$ is the Liouville superoperator acting on variables γ (see general definition (1.4a)), is the set of coordinates and momenta of the test macromolecule in the Heisenberg representation, i.e., at time t.

*Exercise 2.2..*

*Let A(γ)) be a function on phase space. The value A(t) (A in Heisenberg representation) is called* $A(t) \equiv \exp\{i\hat{L}t\} A(\gamma)$. *Prove that* $A(t) = A(\gamma(t))$ *, where* $\gamma(t) \equiv \exp\{i\hat{L}t\}\gamma$ *. The value A(γ)) is assumed to be an analytic function of its arguments.*

The generalized Langevin equation (cf. relation (1.53)) for phase densities has the form:

$$\frac{\partial}{\partial t} f(\Gamma_N;t) = \int d\Gamma'_N i\Omega\left(\Gamma_N \middle| \Gamma'_N\right) f(\Gamma'_N;t) -$$

$$-\int d\Gamma'_N K\left(\Gamma_N \middle| \Gamma'_N;\tau\right) f\left(\Gamma'_N;t-\tau\right) + F^Q\left(\Gamma_N;t\right), \qquad (2.17)$$

where

$$\Omega\left(\Gamma_N \middle| \Gamma'_N\right) = \int \frac{d\Gamma''_N}{\left\langle f\left(\Gamma'_N\right) \middle| f\left(\Gamma''_N\right)\right\rangle} \left\langle f\left(\Gamma'_N\right) \middle| \hat{L}_\gamma f\left(\Gamma_N\right)\right\rangle \qquad (2.17a)$$

- frequency matrix ,

$$F^Q\left(\Gamma_N;t\right) \equiv \exp\left\{i\hat{Q}\hat{L}_\gamma t\right\} i\hat{Q}\hat{L}_\gamma f\left(\Gamma_N\right) \qquad (2.17b)$$

- Langeven generalized force associated with density $f\left(\Gamma_N\right)$, $\hat{Q}_N = 1 - \hat{P}_N$ - projection operator onto a subspace orthogonal to $L_N$,

$$K\left(\Gamma_N \middle| \Gamma'_N;\tau\right) \equiv \int \frac{d\Gamma''_N}{\left\langle f\left(\Gamma'_N\right) \middle| f\left(\Gamma''_N\right)\right\rangle} \left\langle F^Q\left(\Gamma''_N;0\right) \middle| F^Q\left(\Gamma_N;\tau\right)\right\rangle \qquad (2.17c)$$

- memory matrix.

Let's calculate the frequency matrix $\Omega\left(\Gamma_N \mid \Gamma'_N\right)$. For this purpose, we substitute the right-hand side of formula (2.10) into relation (2.17a). Then, using the definition of the Liouville superoperator, we obtain:

$$\Omega\left(\Gamma_N \mid \Gamma'_N\right) = \int d\Gamma''_N \frac{\delta\left(\Gamma'_N - \Gamma''_N\right)}{\rho^*_N\left(\Gamma'_N\right)} \left\langle f\left(\Gamma''_N\right) \middle| \hat{L} f\left(\Gamma_N\right)\right\rangle =$$

$$= \frac{1}{\rho^*_N\left(\Gamma'_N\right)} \left\langle f\left(\Gamma'_N\right) \middle| i\left\{H(\gamma); f\left(\Gamma_N\right)\right\}\right\rangle = \qquad . \qquad (2.18)$$

$$= \frac{i}{\rho^*_N\left(\Gamma'_N\right)} \int d\gamma \rho_{eq}(\gamma) \delta\left(\Gamma'_N - \gamma_N\right) \left\{H(\gamma); f\left(\Gamma_N\right)\right\}$$

The following identity is easy to verify:

$$\left\{\rho_{eq}(\gamma); f\right\} = -\beta \rho_{eq}(\gamma) \left\{H(\gamma); f\right\} , \qquad (2.19)$$

where $f$ is any differentiable function defined on the phase space of the entire system P. To do this, it is sufficient to use the equality $\rho_{eq}(\gamma)=\frac{1}{Z}\exp\{-\beta H(\gamma)\}$ and the definition of Poisson brackets (1.3).

This identity allows us to transform formula (2.18) to the form:

$$\begin{aligned}\Omega\left(\Gamma_N \mid \Gamma'_N\right)&=-\frac{i}{\rho_N^*\left(\Gamma'_N\right)}\int d\gamma\delta\left(\Gamma'_N-\gamma_N\right)\left\{kT\rho_{eq}(\gamma);\delta\left(\Gamma_N-\gamma_N\right)\right\}_\gamma=\\&=-\frac{i}{\rho_N^*\left(\Gamma'_N\right)}\left\{kT\rho_N^*\left(\Gamma'_N\right);\delta\left(\Gamma_N-\Gamma'_N\right)\right\}_{\Gamma'_N}\end{aligned}\quad. \qquad (2.20)$$

where the index in the lower right corner of the Poisson brackets indicates the variables over which differentiation is performed.

Using the definition of an effective single-chain Hamiltonian $H_N^*\left(\Gamma_N\right)$ (see relations (2.5) and (2.6)), equality (2.20) can be transformed to the following form:

$$\Omega\left(\Gamma_N \mid \Gamma'_N\right)=i\left\{H_N^*\left(\Gamma'_N\right);\delta\left(\Gamma_N-\Gamma'_N\right)\right\}_{\Gamma'_N} \quad . \qquad (2.21)$$

The generalized Langevin stochastic force at the initial moment of time is calculated using the general relation (2.17b) for t=0:

$$\begin{aligned}F^Q\left(\Gamma_N\right)&\equiv F^Q\left(\Gamma_N;0\right)=i\hat{Q}_N\hat{L}f\left(\Gamma_N\right)=\\&=i\left(1-\hat{P}_N\right)i\left\{H(\gamma);\delta\left(\Gamma_N-\gamma_N\right)\right\}_\gamma=\\&=\left(\hat{P}_N-1\right)\left\{H(\gamma);\delta\left(\Gamma_N-\gamma_N\right)\right\}_\gamma\end{aligned}\quad . \qquad (2.22)$$

First, let us calculate the projection of the Poisson bracket onto the Liouville space of the test macromolecule using formula (2.12):

$$\begin{aligned}\hat{P}\left\{H(\gamma);\delta\left(\Gamma_N-\gamma_N\right)\right\}_\gamma&=\frac{1}{\rho_N^*\left(\gamma_N\right)}\int d\gamma'\rho^{eq}\left(\gamma'\right)\left\{H\left(\gamma'\right);\delta\left(\Gamma_N-\gamma'_N\right)\right\}_{\gamma'}=\\&=\left\{H_N^*\left(\gamma_N\right);\delta\left(\Gamma_N-\gamma_N\right)\right\}_\gamma\end{aligned}\quad .(2.23)$$

Note that to perform the last transition in formula (2.23), we used identity (2.19) as before. Substituting the ratio (2.23) to the formula (2.22), we get:

$$F^Q(\Gamma_N)=\{\delta(\Gamma_N-\gamma_N);\delta H(\gamma)\}_\gamma\ , \qquad (2.24)$$

where $\delta H(\gamma)\equiv H(\gamma)-H_N^*(\gamma_N)$. (2.24a)

The value $\delta H(\gamma)$ in relation (2.24) is a measure of the fluctuation of energy of intermolecular interactions on the test macromolecule. It is easy to see that all terms of the total Hamiltonian $H(\gamma)$ depending only on the variables of the macromolecule matrix contribute zero to $F^Q(\Gamma_N)$ . For classical systems, there are also no contributions related to kinetic energy. Direct calculations using formula (2.6) lead to the result:

$$F^Q(\Gamma_N)=\sum_n \delta\vec{F}_n^{\mathrm{int}er}\frac{\partial}{\partial\vec{p}_i}\delta(\Gamma_N-\gamma_N), \qquad (2.225)$$

where $\delta\vec{F}_n^{\mathrm{int}er}=-\frac{\partial}{\partial\vec{r}_n}\delta H(\gamma)=-\frac{\partial H}{\partial\vec{r}_n}+\frac{\partial W^*}{\partial\vec{r}_n}$ . (2.25a)

is the fluctuation part of the total intermolecular force acting on the n[th] segment of the test macromolecule.

The stochastic force of Langevin $F^Q(\Gamma_N;t)$ at any given moment in time is determined by the general ratio:

$$\begin{aligned}&F^Q(\Gamma_N;t)\equiv\exp\{i\hat{Q}_N\hat{L}_\gamma t\}F^Q(\Gamma_N)=\\&=\exp\{i\hat{Q}_N\hat{L}_\gamma t\}\{\delta(\Gamma_N-\gamma_N);\delta H(\gamma)\}_\gamma\end{aligned}\ . \qquad (2.26)$$

Substituting this ratio into formula (2.17c), we obtain the following expression for the memory matrix after integration:

$$\begin{aligned}&K(\Gamma_N|\Gamma'_N;t)=\frac{1}{\rho_N^*(\Gamma'_N)}\times\\&\times\left\langle\{\delta(\Gamma'_N-\gamma);\delta H(\gamma)\}_\gamma\middle|\exp\{i\hat{Q}\hat{L}_\gamma t\}\{\delta(\Gamma_N-\gamma);\delta H(\gamma)\}_\gamma\right\rangle_{eq}\end{aligned}\ . \qquad (2.27)$$

## 2.3. Generalized Langevin equation for arbitrary value *a (γ)*.

Let us consider some characteristics of a probe/tagged macromolecule, i.e., some function $a(\gamma_N)$ of its coordinates and momenta. The generalized Langevin equation for phase densities $f(\Gamma_N;t)$ (relation (2.17)) allows us to derive the generalized Langevin equation for the quantity $a(t) \equiv a(\gamma_N(t)) = \exp\{i\hat{L}t\} a(\gamma_N)$.

For this purpose, both parts of equation (2.17) should be multiplied by $a(\Gamma_N)$ and integrated over all $\Gamma_N$.

First, let's perform this operation on the right side of this equation:

$$\begin{aligned} &\int d\Gamma_N \frac{\partial}{\partial t} \delta(\Gamma_N - \gamma_N(t)) a(\Gamma_N) = \\ &= \frac{\partial}{\partial t} \int d\Gamma_N \delta(\Gamma_N - \gamma_N(t)) a(\Gamma_N) = \frac{\partial}{\partial t} a(t) \end{aligned} . \tag{2.28}$$

Similar calculations using formula (2.21) transform the contribution associated with the frequency matrix $\Omega(\Gamma_N \mid \Gamma'_N)$ into the form:

$$\begin{aligned} &\int d\Gamma'_N d\Gamma_N i\Omega(\Gamma_N \mid \Gamma'_N) \delta(\Gamma'_N - \gamma_N(t)) a(\Gamma_N) = \\ &= -\int d\Gamma'_N d\Gamma_N \left\{ H^*_N(\Gamma'_N); \delta(\Gamma_N - \Gamma'_N) \right\}_{\Gamma'_N} a(\Gamma_N) \delta(\Gamma'_N - \gamma_N(t)) = \\ &= -\int d\Gamma'_N \left\{ H^*_N(\Gamma'_N); a(\Gamma'_N) \right\}_{\Gamma'_N} \delta(\Gamma'_N - \gamma_N(t)) = \\ &= \left\{ a(\gamma_N(t)); H^*_N(\gamma_N(t)) \right\}_{\gamma(t)} \end{aligned} . \tag{2.29}$$

Now let's calculate the contribution associated with the memory matrix $K(\Gamma_N \mid \Gamma'_N;\tau)$:

$$\int d\Gamma'_N d\Gamma_N K\left(\Gamma_N \mid \Gamma'_N;\tau\right)\delta\left(\Gamma'_N-\gamma_N(t-\tau)\right)a\left(\Gamma_N\right)$$

$$=\int\frac{d\Gamma'_N d\Gamma_N}{\rho^*_N\left(\Gamma'_N\right)}\left\langle\left\{\delta\left(\Gamma'_N-\gamma_N(t)\right);\delta H\left(\gamma_N\right)\right\}_\gamma \mid e^{i\hat{Q}\hat{L}_\gamma\tau}\left\{\delta\left(\Gamma_N-\gamma_N\right);\delta H\left(\gamma_N\right)\right\}_\gamma\right\rangle$$

$$\times a\left(\Gamma_N\right)= \qquad .(2.30)$$

$$=\frac{1}{\rho^*_N\left(\gamma_N(t-\tau)\right)}\left\langle\left\{\delta\left(\gamma_N(t-\tau)-\gamma_N\right);\delta H\left(\gamma_N\right)\right\}_\gamma \mid e^{i\hat{Q}\hat{L}_\gamma\tau}\left\{a\left(\gamma_N\right);\delta H\left(\gamma\right)\right\}_\gamma\right\rangle$$

Finally, let us calculate the contribution from the last term in relation (2.17), the so-called stochastic force associated with the quantity $a(\gamma_N)$:

$$F^Q_a(t)\equiv\int d\Gamma_N F^Q\left(\Gamma_N;t\right)a\left(\Gamma_N\right)=$$

$$=\int d\Gamma_N \exp\left\{i\hat{Q}\hat{L}_\gamma t\right\}\left\{\delta\left(\Gamma_N-\gamma_N\right);\delta H\left(\gamma\right)\right\}_\gamma a\left(\Gamma_N\right)= \quad . \qquad (2.31)$$

$$=\exp\left\{i\hat{Q}\hat{L}_\gamma t\right\}\left\{a\left(\gamma_N\right);\delta H\left(\gamma\right)\right\}_\gamma$$

Combining equations (2.28) – (2.31), we obtain the following kinetic equation for $a(t)$:

$$\frac{d}{dt}a(t)=\left\{a\left(\gamma(t)\right);H^*_N\left(\gamma_N(t)\right)\right\}-$$

$$-\int_0^t\frac{d\tau}{\rho^*_N\left(\gamma_N(t-\tau)\right)}\left\langle\left\{\delta\left(\gamma_N(t-\tau)-\gamma_N\right);\delta H\left(\gamma\right)\right\}_\gamma \mid \exp\left\{i\hat{Q}\hat{L}_\gamma\tau\right\}\left\{a(\gamma_N);\delta H\left(\gamma\right)\right\}_\gamma\right\rangle_{eq}$$

$$+\exp\left\{i\hat{Q}\hat{L}_\gamma t\right\}\left\{a\left(\gamma_N\right);\delta H\left(\gamma\right)\right\}. \qquad (2.32)$$

The quantity $\delta H(\gamma)$ does not depend on the momentum of the test macromolecule (see relation (2.24)), while the Dirac δ-function depends on the difference between coordinates. These circumstances allow us to rewrite equation (2.32) as follows:

$$\frac{d}{dt}a(t)=\left\{a\left(\gamma(t)\right);H^*_N\left(\gamma_N(t)\right)\right\}+\exp\left\{i\hat{Q}\hat{L}t\right\}\left\{a;\delta H\right\}+$$

$$+\int_0^t\frac{d\tau}{\rho^*_N\left(t-\tau\right)}\sum_n\frac{\partial}{\partial\vec{p}_n(t-\tau)}\left\langle\delta\vec{F}^{\mathrm{inter}}_n\delta\left(\gamma_N(t-\tau)-\gamma_N\right)\exp\left\{i\hat{Q}\hat{L}\tau\right\}\left\{a;\delta H\right\}_\gamma\right\rangle.(2.33)$$

## 2.4. Generalized Langevin equation for radius vectors and momenta of segments of a test macromolecule.

Equation (2.33) is an exact consequence of the equations of motion for the entire system and is valid for any differentiable function $a\left(\left\{\vec{r}_1,\ldots\vec{r}_N;\vec{p}_1,\ldots\vec{p}_N\right\}\right)$ of the test macromolecule. In particular, this also applies

$$a=\vec{r}_k, \qquad (2.34)$$

where k=1,2,…,N.

As already noted, the quantity $\delta H(\gamma)$ for classical systems does not depend on the momentum of the test macromolecule (see relations (2.24a) and (2.5), (2.6)). Therefore,

$$\left\{\vec{r}_k;\delta H(\gamma)\right\}=0 \quad . \qquad (2.35)$$

It follows that for case (2.34), equation (2.33) leads to a “trivial” result:

$$\frac{d}{dt}\vec{r}_k(t)=\frac{1}{m}\vec{p}_c(t) \quad . \qquad (2.36)$$

More interesting is the case when

$$a=\vec{p}_k. \qquad (2.37)$$

The first term on the right-hand side of equation (2.33) is equal to the effective intramolecular force acting on segment k from all other segments:

$$\vec{F}_k^* \equiv \vec{F}_k^{\mathrm{int}ra} \equiv \left\{\vec{p}_k;\delta H_N^*(\gamma_N)\right\}=-\frac{\partial}{\partial\vec{r}_k}W^*\left\{\vec{r}_i\right\} \quad . \qquad (2.38)$$

Usually, the force $\vec{F}_k^*\left\{\vec{r}_i\right\}$, which in a general complex case is a functional of the conformation of the test macromolecule, is called the mean force acting on segment k. Note that the latter name is more appropriate, since the force $\vec{F}_k^*$ actually contains contributions from both intramolecular and intermolecular interactions. The contribution from intermolecular interactions is averaged over all distributions of matrix macromolecules at a fixed conformation of the test macromolecule.

The second term in expression (2.33) is the stochastic component of the total intermolecular force acting on segment k:

$$\vec{F}_n^Q(t) \equiv \exp\left\{i\hat{Q}_N\hat{L}t\right\}\left\{\vec{p}_n;\delta H(\gamma)\right\} = \exp\left\{i\hat{Q}_N\hat{L}t\right\}\delta\vec{F}_n^{\text{inter}} \quad . \tag{2.39}$$

It is useful to emphasize once again that $\delta\vec{F}_n^{\text{inter}}$, determined by the ratio (2.25a), is the fluctuation part of the total intermolecular force acting on segment n at time t=0, with a fixed conformation of the test macromolecule. By direct calculation, it is easy to verify the following equality:

$$\vec{F}_k^Q(t) \equiv \exp\left\{i\hat{Q}_N\hat{L}\hat{Q}_N t\right\}\hat{Q}_N\vec{F}_k = \exp\left\{i\hat{Q}_N\hat{L}\hat{Q}_N t\right\}\hat{Q}_N\vec{F}_k^{\text{inter}}, \tag{2.40}$$

where $\vec{F}_k$ - total force acting on segment k at t=0.

*Exercise 2.3.*

*Obtain relation (2.40). To do this, expand the exponential operator into a Taylor series. Use the definition $\hat{Q}_N = 1 - \hat{P}_N$ and relation (2.13).*

From equation (2.40), it can be seen that the generalized stochastic force $\vec{F}_k^Q(t)$ acting on segment k evolves over time under the action of the evolution superoperator:

$$\hat{S}^Q(t) \equiv \exp\left\{i\hat{Q}_N\hat{L}\hat{Q}_N t\right\}, \tag{2.41}$$

sometimes referred to as the projective evolution operator. Since the superoperators $\hat{Q}$ and $\hat{L}$ are Hermitian, the projection evolution operator is unitary. Recall that the real evolution superoperator, or propagator, which determines the actual movements of particles in space, is given by the relation:

$$\hat{S}(t) \equiv \exp\left\{i\hat{L}t\right\} \quad . \tag{2.42}$$

Let us now proceed to calculate the integral contribution to equation (2.33) for the case $a = \vec{r}_n$ we are interested in. The presence of the Dirac δ-function and the Gibbs equilibrium distribution function allows us to transform this term as follows:

$$\int_0^t \frac{d\tau}{\rho_N^*(t-\tau)} \sum_k \frac{\partial}{\partial p_k^\alpha(t-\tau)} \left\langle \delta \vec{F}_k^{Q\alpha} \delta\big(\gamma_N(t-\tau)-\gamma_N\big) \vec{F}_n^Q(\tau) \right\rangle =$$

$$= -\frac{1}{kT} \int_0^t \frac{d\tau}{\rho_N^*(t-\tau)} \sum_k V_k^\alpha \left\langle \delta \vec{F}_k^{Q\alpha} \delta\big(\gamma_N(t-\tau)-\gamma_N\big) \vec{F}_n^Q(\tau) \right\rangle + , \qquad (2.43)$$

$$+\int_0^t \frac{d\tau}{\rho_N^*(t-\tau)} \sum_k \left\langle \delta \vec{F}_k^{Q\alpha} \delta\big(\gamma_N(t-\tau)-\gamma_N\big) \frac{\partial}{\partial p_k^\alpha} \vec{F}_n^Q(\tau) \right\rangle$$

where a=x,y,z is an index denoting the corresponding coordinate of the vector, and the repeating index implies summation, $\vec{V}_k(t)$ is the velocity vector of segment number k at time t.

The standard approximation consists in neglecting the second term on the right-hand side of equation (2.43), i.e., the term $\int_0^t \frac{d\tau}{\rho_N^*(t-\tau)} \sum_k \left\langle \delta \vec{F}_k^{Q\alpha} \delta\big(\gamma_N(t-\tau)-\gamma_N\big) \frac{\partial}{\partial p_k^\alpha} \vec{F}_n^Q(\tau) \right\rangle$. The limits of the validity of this approximation, as far as we know, have not yet been studied in sufficient detail. It is relatively easy to show that for a system of harmonic oscillators, this term is equal to zero. Consequently, it depends directly on the anharmonicity of the interaction potential between the particles of the system. We also note that $\frac{\partial}{\partial p_k^\alpha} \vec{F}_n^Q(\tau=0)$.From physical considerations, it seems natural that if dependence on the initial momentum $p_k^\alpha$ of the Langevin stochastic force $\vec{F}_n^Q(\tau)$ appears at a later time $\tau$, it should quickly decay again at times larger than the characteristic collision time. In very general terms, it can be noted that such terms in equation (2.43) would imply the emergence of friction forces that are nonlinear with respect to the velocity of the test macromolecule.

Relationships (2.36), (2.38) (2.39), and (2.43), in which the second term on the right-hand side is omitted, lead to the Generalized Langevin Equation, which describes the evolution of the coordinates and momenta of the segments of the test macromolecule:

$$\frac{d}{dt}\vec{p}_n(t) = -\frac{\partial}{\partial \vec{r}_n}\tilde{W}^*\left(\{\vec{r}_i(t)\}\right) - \sum_k \int_0^t d\tau \Gamma_{nk}^{\alpha\beta}(\tau;t-\tau)V_k^{\beta}(t-\tau)\vec{e}_\alpha + \vec{F}_n^Q(t)$$

$$\frac{d}{dt}\vec{r}_n(t) \equiv \vec{V}_n(t) = \frac{1}{m}\vec{p}_n(t) \quad , \qquad (2.44)$$

where $\vec{e}_\alpha$ – unit vector directed along the axis α,

$$\Gamma_{nk}^{\alpha\beta}(\tau;t-\tau) = \frac{1}{kT\rho_N^*(t-\tau)}\left\langle F_k^{Q\beta}(\gamma)\delta\left(\gamma_N - \gamma_N(t-\tau)\right)F_n^{Q\alpha}(\tau)\right\rangle, \qquad (2.44a)$$

– memory matrix.

The generalized Langevin equation is a system of 6N nonlinear integro-differential relations (2.44). In order for these exact relations to actually become equations, it is necessary to know the memory matrix explicitly. $\Gamma_{nk}^{\alpha\beta}(\tau;t-\tau)$. Equation (2.44a) shows that the memory matrix $\Gamma_{nk}^{\alpha\beta}(\tau;t-\tau)$ is, in general, a nonlocal second-order tensor.

**2.5. Projection propagator, projection dynamics.**

The time evolution, or time dynamics, of the generalized stochastic force of Langevin $\vec{F}_n^Q(t)$ acting on the nth segment of the test chain is determined by the projection propagator $\hat{S}^Q(t)$ given by relation (2.41).

Let us consider a certain physical quantity $a(\gamma)$. Using a projection propagator, we can construct a new quantity:

$$a^Q(t) \equiv \exp\left\{i\hat{Q}_N\hat{L}\hat{Q}_N\right\}a(\gamma). \qquad (2.45)$$

Relationships (2.45) $\forall\, a(\gamma)$ determine the so-called projection dynamics, and the value can be interpreted as the value $a^Q(t)$ of the physical quantity a at time t as a result of projection evolution. This value, generally speaking, differs from the actual value of a at time t, defined as:

$$a(t) \equiv \exp\left\{i\hat{L}t\right\}a. \qquad (2.46)$$

Let's examine some general properties of projection dynamics.

Let us begin by considering the projection evolution of the position vectors of the test chain segments. By definition

$$\vec{r}_n^{\,Q}(t) \equiv \exp\left\{ i\hat{Q}_N \hat{L} \hat{Q}_N t \right\} \vec{r}_n \ . \tag{2.47}$$

Differentiating both sides of this equation with respect to time, we obtain:

$$\frac{d}{dt}\vec{r}_n^{\,Q}(t) = \exp\left\{ i\hat{Q}_N \hat{L} \hat{Q}_N t \right\} i\hat{Q}_N \hat{L} \hat{Q}_N \vec{r}_n . \tag{2.48}$$

But

$$\hat{Q}_N \vec{r}_n = (1 - \hat{P}_N)\vec{r}_n = \vec{r}_n - \vec{r}_n = 0 , \tag{2.49}$$

therefore

$$\frac{d}{dt}\vec{r}_n^{\,Q}(t) = 0 . \tag{2.50}$$

Equality (2.50) means that as a result of projection evolution, the initial conformation of the test chain does not change.

Similar calculations show that the momentum of the probe chain segments also does not change over time as a result of projection evolution:

$$\frac{d}{dt}\vec{p}_n^{\,Q}(t) = \exp\left\{ i\hat{Q}_N \hat{L} \hat{Q}_N t \right\} i\hat{Q}_N \hat{L} \hat{Q}_N \vec{p}_n = 0 . \tag{2.51}$$

Let us now consider the projection evolution of the position vector of the n-th segment of the μ-th macromolecule of the matrix:

$$\vec{r}_{\mu n}^{\,Q}(t) \equiv \exp\left\{ i\hat{Q}_N \hat{L} \hat{Q}_N t \right\} \vec{r}_{\mu n} \ . \tag{2.52}$$

Differentiating both sides of equation (2.52) with respect to time, we obtain

$$\frac{d}{dt}\vec{r}_{\mu n}^{\,Q}(t) = \exp\left\{ i\hat{Q}_N \hat{L} \hat{Q}_N t \right\} \hat{Q}_N i\hat{L} \hat{Q}_N \vec{r}_{\mu n} . \tag{2.53}$$

But now, unlike (2.49):

$$\hat{Q}_N \vec{r}_{\mu n}^{\,Q} = \vec{r}_{\mu n}^{\,Q} , \tag{2.54}$$

And

$$i\hat{L}\vec{r}_{\mu n} = i^2 \left\{ H(\gamma); \vec{r}_{\mu n} \right\} = \vec{V}_{\mu n} , \tag{2.55}$$

where $\vec{V}_{\mu n}$ – the velocity of the n-th segment of the μ-th matrix chain in Schrödinger's representation. It follows from this that:

$$\frac{d}{dt}\vec{r}^{\,Q}_{\mu n}(t) = \vec{V}^{Q}_{\mu n}(t) = \frac{1}{m}\vec{p}^{\,Q}_{\mu n}(t) \ , \qquad (2.56)$$

i.e., the time derivative of the projection value of the radius vectors of the matrix segments coincides with the projection values of the velocities.

The projection evolution of matrix chain segment pulses is determined by the relation:

$$\vec{p}^{\,Q}_{\mu n}(t) \equiv \exp\left\{i\hat{Q}_N\hat{L}\hat{Q}_N t\right\}\vec{p}_{\mu n} \ . \qquad (2.57)$$

Differentiating both sides of this equation gives:

$$\frac{d}{dt}\vec{p}^{\,Q}_{\mu n}(t) = \exp\left\{i\hat{Q}_N\hat{L}\hat{Q}_N t\right\}\hat{Q}_N i\hat{L}\hat{Q}_N\vec{p}_{\mu n} \quad . \qquad (2.58)$$

Then

$$\hat{Q}_N\vec{p}_{\mu n} = \vec{p}_{\mu n} \qquad (2.59)$$

$$i\hat{L}\vec{p}_{\mu n} = \left\{\vec{p}_{\mu n}; H(\gamma)\right\} = \vec{F}_{\mu n}, \qquad (2.60)$$

where $\vec{F}_{\mu n}$ - total force acting on the n-th segment of the μ-th chain of the matrix.

Consider the quanity:

$$\hat{Q}_N\vec{F}_{\mu n} = \vec{F}_{\mu n} - \hat{P}_N\vec{F}_{\mu n}. \qquad (2.61)$$

The second term on the right-hand side of equation (2.56) contains the projection operator $\hat{P}_N$ onto the Liouville space of the test macromolecule. Therefore, only terms associated with interactions between the n-th segment of the μ-th matrix and segments of the test macromolecule contribute to the nonzero value $\hat{P}_N\vec{F}_{\mu n}$. Using ratio (2.13), it is easy to verify the validity of the following estimate:

$$\hat{P}_N \vec{F}_{\mu n} \propto \frac{Nb^3 \hat{P}_N}{V}, \qquad (2.62)$$

where V – total volume og all system, b – length of Kugh segment, N – number of Kugh segments per polymer chain.

It is clear that in the thermodynamic limit V→∞ at C=const, C is the concentration of macromolecules:

$$\hat{P}_N \vec{F}_{\mu n} = 0. \qquad (2.63)$$

Based on relations (2.54)-(2.58), we obtain the differential equation

$$\frac{d}{dt}\vec{P}^{Q}_{\mu n}(t) = \vec{F}^{Q}_{\mu n}(t), \qquad (2.64)$$

where $\vec{F}^{Q}_{\mu n}(t) \equiv \exp\left\{ i\hat{Q}_N \hat{L} \hat{Q}_N t \right\} \vec{F}_{\mu n}$.

Summarizing the results obtained, we have the following:

$$\begin{aligned}
&\frac{d}{dt}\vec{r}^{Q}_{n}(t) = 0 \\
&\frac{d}{dt}\vec{p}^{Q}_{n}(t) = 0 \\
&\frac{d}{dt}\vec{r}^{Q}_{\mu n}(t) = \frac{1}{m}\vec{p}^{Q}_{\mu n}(t) \\
&\frac{d}{dt}\vec{p}^{Q}_{\mu n}(t) = \vec{F}^{Q}_{\mu n}(t)
\end{aligned} . \qquad (2.65)$$

This is all the exact results known to the author. To obtain further results, it is necessary to introduce various approximations based, as a rule, on certain physical reasons and which, from a theoretical point of view, are uncontrollable approximations. The success of such approximations is measured by the success of their predictions when compared with experimental data. From a theoretical point of view, the controllability of approximations means that either:

1. An exact solution to the problem is presented.
2. The problem has a small parameter that allows us to construct a perturbation theory based on the known exact solution so that we can systematically evaluate the accuracy of approximations. In condensed matter physics, only a small number of problems allow for controlled approximations in theoretical considerations. This is closely related to the problem of closing the infinite chain of BBGKY (Bogoliubov-Born-Green-Kirckwood-Yvon) equations, discussed in the first chapter.

Before discussing the simplest approximations, let us note the structure of the memory matrix (2.44a):

$$\Gamma_{nk}^{\alpha\beta}(\tau;t-\tau)=\frac{1}{kT\rho_N^*(t-\tau)}\left\langle F_k^{Q\beta}(\gamma)\delta\left(\gamma_N-\gamma_N(t-\tau)\right)F_n^{Q\alpha}(\tau)\right\rangle$$

$$=\frac{1}{kT}\left\langle F_k^{Q\beta}(0)F_n^{Q\alpha}(\tau)\right\rangle_{\gamma_{N(t-\tau)}}^* \qquad (2.66)$$

From equation (2.66) it can be seen that the memory matrix has the structure of a conditional mean, with the initial condition that the test macromolecule has a conformation that coincides with its actual conformation $\gamma_N\left(t-\tau\right)$ at time $t-\tau$ and further, over time, the stochastic Langevin force develops under the action of the projection operator of evolution $\exp\left\{i\hat{Q}_N\hat{L}\hat{Q}_N\tau\right\}$.

## 2.6. Average strength potential and memory matrix. Rouse model as the simplest approximation.

The potential of the mean force $\tilde{W}^*\left(\left\{\vec{r}_i(t)\right\}\right)$ in equation (2.44) is determined, in principle, by methods of equilibrium statistical mechanics. In polymer melts, as is well known (see, for example, [15-16] and the literature cited therein), the distribution of coarse-grained conformations of macromolecules is approximated with good accuracy by a Gaussian distribution. In that case, the average power potential, accurate to within an insignificant constant term, is equal to:

$$\tilde{W}^*\left(\left\{\vec{r}_i(t)\right\}\right)=\frac{3kT}{2b^2}\sum_{n=1}^{N}\left(\vec{r}_n-\vec{r}_{n-1}\right)^2, \qquad (2.67)$$

where $\vec{r}_n$- position-vector of the n-th segment of a probe macromolecule радиус-вектор n-го сегмента макромолекулы, $b$- Kuhn segment length a. Substituting this expression into the generalized Langevin equation (2.44), we obtain the effective intramolecular entropic force of polymer chain tension acting on the nth segment of the macromolecule appearing in the Rouse model ). With analytical approximations for the memory matrix, the situation is significantly more complex. The memory matrix $\Gamma_{nk}^{\alpha\beta}(\tau;t-\tau)$ in principle contains all dynamic information related to intermolecular interactions, in particular all entanglement effects. The fluctuating part of the intermolecular force acting on the n th segment of the probe macromolecule can be expressed through the fluctuation of the polymer matrix density:

$$\vec{F}_n^Q(\tau)=\int d^3\vec{r}\vec{f}(\vec{r})\delta\rho_n^Q\left(\vec{r};\tau;\gamma_N(t-\tau)\right), \qquad (2.68)$$

where $\vec{f}(\vec{r})=-\frac{\partial}{\partial\vec{r}}U(\vec{r})$- the force with which the polymer segment of the matrix, distanced from the segment of the test macromolecule under consideration by a radius vector $\vec{r}$, acts on the n-th segment of the test macromolecule under consideration. The dynamic evolution of matrix density fluctuations develops in accordance with projection dynamics:

$$\delta\rho_n^Q\left(\vec{r};\tau;\gamma_N(t-\tau)\right)=\exp\left\{i\hat{Q}\hat{L}\hat{Q}\tau\right\}\left(\rho_n\left(\vec{r};\tau;\gamma_N(t-\tau)\right)-\rho_n^*\left(\vec{r};\gamma_N(t-\tau)\right)\right), \qquad (2.69)$$

where $\rho_n\left(\vec{r};\tau;\gamma_N(t-\tau)\right)$- the concentration of matrix segments at a given moment in time $t-\tau$ with a fixed configuration $\gamma_N(t-\tau)$ of probe macromolecule, $\rho_n^*\left(\vec{r};\gamma_N(t-\tau)\right)$- average concentration of matrix segments, assuming that the test chain is frozen in conformation $\gamma_N(t-\tau)$.

Then the memory matrix can be written as follows:

$$\Gamma_{nk}^{\alpha\beta}(\tau;t-\tau)=\frac{1}{kT}\int d^3\vec{r}d^3\vec{r}\,'f_n^{\alpha}(\vec{r})f_k^{\beta}(\vec{r}\,')\left\langle\delta\rho_k^Q(0;\vec{r})\delta\rho_n^Q(\tau;\vec{r}\,')\right\rangle_{\gamma_N(t-\tau)}^{*}. \quad (2.70)$$

The dynamic correlation function $\left\langle\delta\rho_k^Q(0;\vec{r})\delta\rho_n^Q(\tau;\vec{r}\,')\right\rangle_{\gamma_N(t-\tau)}^{*}$ is determined by the projection dynamics and is a conditional average and, generally speaking, is a complex functional of the course graned conformation $\gamma_N(t-\tau)$ of the test macromolecule at the time moment $t-\tau$. The latter circumstance does not exclude the importance of reptational mobility modes in the dynamics of macromolecules. In isotropic dynamic models, the simplest of which is the Rouse model, when interpreting the correlation function under discussion, the conditional mean is approximated by the equilibrium mean, i.e. pre-averaging is performed:

$$\left\langle\delta\rho_k^Q(0;\vec{r})\delta\rho_n^Q(\tau;\vec{r}\,')\right\rangle_{\gamma_N(t-\tau)}^{*}\approx\left\langle\delta\rho_k^Q(0;\vec{r})\delta\rho_n^Q(\tau;\vec{r}\,')\right\rangle_{eq}. \quad (2.71)$$

After this approximation, the projection dynamic correlation function $\left\langle\delta\rho_k^Q(0;\vec{r})\delta\rho_n^Q(\tau;\vec{r}\,')\right\rangle_{\gamma_N(t-\tau)}^{*}$ ceases to be a functional of the macromolecule conformation at the time $t-\tau$ . Since polymer melts of flexible-chain macromolecules are isotropic, the memory matrix becomes a scalar with respect to spatial rotations:

$$\Gamma_{nk}^{\alpha\beta}(\tau;t-\tau)=\frac{1}{3kT}\left\langle\vec{F}_k^Q(0)\vec{F}_n^Q(\tau)\right\rangle_{eq}\delta_{\alpha\beta} \ . \quad (2.73)$$

Furthermore, generally speaking, the memory matrix is non-local, i.e. it takes into account long-range effects similar to hydrodynamic ones that arise between different segments of the test macromolecule as a result of interaction with the matrix.

The next simplest approximation is to neglect the mentioned non-locality:

$$\Gamma_{nk}^{\alpha\beta}(\tau;t-\tau)=\frac{1}{3kT}\left\langle\vec{F}_n^Q(0)\vec{F}_n^Q(\tau)\right\rangle_{eq}\delta_{\alpha\beta}\delta_{nk}. \quad (2.74)$$

If we consider the memory function as a rapidly changing variable:

$$\Gamma_{nk}^{\alpha\beta}(\tau;t-\tau)=2\zeta\delta_{\alpha\beta}\delta_{nk}\delta(\tau), \tag{2.75}$$

where

$$\zeta=\frac{1}{3kT}\int_0^\infty d\tau\left\langle \vec{F}_n^Q(0)\vec{F}_n^Q(\tau)\right\rangle_{eq} \tag{2.76}$$

is local friction coefficient, equations (2.44) transform into the classical Rouse equations:

$$m\ddot{\vec{r}}_n(t)=\frac{3k_BT}{b^2}\{\vec{r}_{n+1}(t)+\vec{r}_{n-1}(t)-2\vec{r}_n(t)\}-\zeta\dot{\vec{r}}_n(t)+\vec{F}_n^L(t). \tag{2.77}$$

The reader can find a discussion of other approximations, similar to renormalized Rouse models and polymer coupled mode models, in [17-20]. A discussion of the effectiveness of the Rouse model in describing the dynamic properties of polymer melts with molecular masses below critical molecular masses can be found in [21]. Improving the aforementioned approximations is proposed as a topic for further research. A familiarity with [17-21] is useful for general development.

## 2.7. Projection dynamics for a system of harmonic oscillators. Exact result.

Let's consider a system of coupled oscillators. We'll call the oscillators selected from among them, based on some principle, a macromolecule, and all the others, a matrix. It is obvious that the general formalism developed above is also applicable to this situation. Let us show that in this case the projection dynamics of the quantities determining the evolution of the memory matrix satisfies Hamilton's equations with the Hamiltonian $\delta H(\gamma)=H(\gamma)-H^*(\gamma_N)$ defined by relations (2.6) and (1.24a). In other words, the projection dynamics corresponding to the projection operator (2.11) coincides with the real dynamics of the matrix, provided that the conformation of the test chain is fixed. The Liouville superoperator of the entire system can be written as a sum:

$$\hat{L}_{\gamma} = \hat{L}_{\mu} + \hat{L}_{N}, \tag{2.78}$$

where $\hat{L}_{\mu}$ - part of the full Liouville operator acting on the dynamic variables of the particles of the matrix, a $\hat{L}_{N}$ - part of the full Liouville superoperator acting on the dynamic variables of the probe chain. Note that the following identity holds:

$$\hat{L}_{\mu} = \hat{Q}_{N}\hat{L}_{\mu}\hat{Q}_{N}. \tag{2.79}$$

Indeed, $\hat{Q}_{N} = 1 - \hat{P}_{N}$, $\hat{L}_{\mu}\hat{P}_{N} = 0$ since is linear in the operations of differentiation with respect to the dynamic variables of the matrix particles, and any value $\hat{P}_{N}A$ in accordance with relation (2.14) depends only on the dynamic variables of the test particle. The quantity $\hat{P}_{N}\hat{L}_{\mu} = 0$ since, as is easy to verify, $\hat{L}_{\mu}\delta H(\gamma) = 0$ , and the action of the projection operator $\hat{L}_{\mu}\delta H(\gamma) = 0$, as follows from the already mentioned relation (2.14), is $\hat{P}_{N}$ a Gibbs averaging with the Hamiltonian $\delta H(\gamma) = H(\gamma) - H^{*}(\gamma_{N})$.

Let us now return to expression (2.66) for the memory matrix:

$$\begin{aligned}\Gamma_{nk}^{\alpha\beta}(\tau;t-\tau) &= \frac{1}{kT\rho_{N}^{*}(t-\tau)}\left\langle F_{k}^{Q\beta}(\gamma)\delta\left(\gamma_{N} - \gamma_{N}(t-\tau)\right)F_{n}^{Q\alpha}(\tau)\right\rangle \\ &= \frac{1}{kT}\left\langle F_{k}^{Q\beta}(0)F_{n}^{Q\alpha}(\tau)\right\rangle_{\gamma_{N(t-\tau)}}^{*}\end{aligned}. \tag{2.80}$$

The dynamics of the fluctuation part of the force acting on the n-th particle of a test macromolecule is determined by the projection propagator:

$$F_{n}^{Q\alpha}(\tau) = \exp\left\{i\tau\hat{Q}_{N}\hat{L}\hat{Q}_{N}\right\}F_{n}^{Q\alpha}, \tag{2.81}$$

This propagator, in accordance with relation (2.79), can be represented as:

$$F_{n}^{Q\alpha}(\tau) = \exp\left\{i\tau\left(\hat{L}_{\mu} + \hat{Q}_{N}\hat{L}_{N}\hat{Q}_{N}\right)\right\}F_{n}^{Q\alpha}. \tag{2.82}$$

Next, relation (2.82) can be transformed as follows:

$$F_n^{Q\alpha}(\tau)=\exp\left\{i\tau\hat{L}_\mu\right\}\hat{T}\exp\left\{i\int_0^\tau d\tau_1\hat{Q}_N\hat{L}_N\left(\tau_1\right)\hat{Q}_N\right\}F_n^{Q\alpha}, \tag{2.83}$$

where $\hat{T}\exp\left\{i\int_0^\tau d\tau_1\hat{Q}_N\hat{L}_N\left(\tau_1\right)\hat{Q}_N\right\}$ - the chronological Dyson exponent, and the time dependence of the coordinates and momenta of the particles of the matrix in the superoperator $\hat{L}_N\left(\tau_1\right)$ is determined by the dynamic evolution of the matrix with the Hamiltonian $\delta H(\gamma)=H(\gamma)-H^*(\gamma_N)$. Let us consider the Liouville superoperator of a probe chain:

$$\hat{L}_N\left(\tau_1\right)=i\sum_n\left(\left(\vec{F}_n^{\text{int}ra}\{\vec{r}_i\}+\sum_\mu\vec{F}_{\mu n}\left(\vec{r}_\mu\left(\tau_1\right)-\vec{r}_n\right)\right)\frac{\partial}{\partial\vec{p}_n}+\frac{\vec{p}_n}{m_n}\frac{\partial}{\partial\vec{r}_n}\right), \tag{2.84}$$

where the summation over $\mu$ is carried out over all particles of the matrix, the summation over is carried out $n$ over the particles of the probe chain, $\vec{F}_{\mu n}\left(\vec{r}_\mu\left(\tau_1\right)-\vec{r}_n\right)$ denotes the force with which the particle of the matrix with number $\mu$ is acting at a given moment in time $\tau_1$ on particle of the test chain with number $n$ , $\vec{r}_\mu\left(\tau_1\right)=\exp\left\{i\tau_1\hat{L}_\mu\right\}\vec{r}_\mu$ . In the case of a system of harmonic oscillators, the forces acting between particles are linear in coordinates; in addition, at the initial moment of time they do not depend on the initial momenta of the particles of the test chain. Therefore

$$\begin{aligned}&\hat{Q}_N\hat{L}_N\left(\tau_1\right)\hat{Q}_NF_n^{Q\alpha}\\&=i\hat{Q}_N\sum_n\left(\left(\vec{F}_n^{\text{int}ra}\{\vec{r}_i\}+\sum_\mu\vec{F}_{\mu n}\left(\vec{r}_\mu\left(\tau_1\right)-\vec{r}_n\right)\right)\frac{\partial}{\partial\vec{p}_n}+\frac{\vec{p}_n}{m_n}\frac{\partial}{\partial\vec{r}_n}\right)\hat{Q}_NF_n^{Q\alpha}=,\\&=i\hat{Q}_N\sum_n\frac{\vec{p}_n}{m_n}\frac{\partial}{\partial\vec{r}_n}\hat{Q}_NF_n^{Q\alpha}=i\left(1-\hat{P}_N\right)\sum_n\frac{\vec{p}_n}{m_n}\vec{A}_n\end{aligned} \tag{2.85}$$

where $\vec{A}_n$ is a second-order tensor independent of coordinates and momenta of matrix paricles, determined by the elastic constants of the system. For this reason it is obvious that

$$\left(1-\hat{P}_N\right)\sum_n\frac{\vec{p}_n}{m_n}\vec{A}_n=0. \tag{2.86}$$

From this it is clear that in relation (2.83) the chronological Dyson exponent can be set equal to unity, and the projection dynamics in the memory matrix (2.80) turns out to be Hamiltonian dynamics with Hamiltonian $\delta H(\gamma) = H(\gamma) - H^*(\gamma_N)$.

*. Exercise 2.4. Reproduce the omitted calculations.*

*To illustrate the results of this section, consider a one-dimensional oscillator formed by two identical particles. The Hamiltonian of this particle is:*

$$H = \frac{p_1^2}{2m} + \frac{p_2^2}{2m} + \frac{\kappa}{2}\left(x_2 - x_1\right)^2. \qquad (2.87)$$

*We will formally consider particle number 1 as a test particle, and particle number 2 as a matrix. Note that the formalism developed above is general and can be applied to this exactly solvable case. The generalized Langevin equation (2.44) for this simplest case has the following form:*

$$\frac{d}{dt}p_1(t) = -\frac{1}{kTm}\int_0^t d\tau \left\langle F_{21}^Q(\tau) F_{21}^Q(0)\right\rangle_{x_1;p_1} p_1(t-\tau) + \vec{F}_{21}^Q(t). \qquad (2.88)$$

*From the theorem proved above it follows*

$$\left\langle F_{21}^Q(\tau) F_{21}^Q(0)\right\rangle_{x_1;p_1} = \left\langle \left(F_{21}^Q(0)\right)^2\right\rangle_{x_1;p} \cos\left(\sqrt{\frac{\kappa}{m}}t\right). \qquad (2.89)$$

*In our case, the average* $\left\langle \left(F_{21}^Q(0)\right)^2\right\rangle_{x_1;p} = \kappa kT$ *does not depend on the initial coordinate and momentum of the first particle. We multiply both sides of equation (2.88) by the initial value of the momentum of the first particle and average over all its realizations. We obtain the following equation for the momentum-momentum autocorrelation function:*

$$\frac{d}{dt}\left\langle p_1(t) p_1(0)\right\rangle = -\frac{\kappa}{m}\int_0^t d\tau \cos\left(\sqrt{\frac{\kappa}{m}}\tau\right)\left\langle p_1(t-\tau) p_1(0)\right\rangle. \qquad (2.90)$$

*This integro-differential equation can be solved exactly. To do this, it's necessary to perform the Laplace transform. Solve the resulting elementary*

*algebraic equation. Then perform the inverse Laplace transform. The result is as follows:*

$$\langle p_1(t)\,p_1(0)\rangle = \frac{\langle p_1^2\rangle}{2}\left(1+\cos\left(\sqrt{\frac{2\kappa}{m}}t\right)\right) = \frac{mkT}{2}\left(1+\cos\left(\sqrt{\frac{2\kappa}{m}}t\right)\right). \qquad (2.91)$$

*The time-independent term in our simplified model corresponds to the displacement of the center of mass of the two-particle system, and the oscillating term corresponds to the harmonic oscillations of an oscillator formed from two identical particles.*

*Exercise 2.5. Show that for a system of harmonic oscillators the transition from relation (2.43) to (2.44) is exact.*

**References**

1. Zwanzig R. J. Chem. Phys. **33**, p.1338 (1960);
2. Zwanzig R. Phys. Rev. **124**, p.985 (1961);
3. Mori H. Progr. Theor. Phys. **34**, p.765 (1965);
4. Mori H. Progr. Theor. Phys. **33**, p.423 (1965);
5. Zwanzig R. Ann. Rev. Phys. Chem. **16**, p.67 (1965);
6. Forster D. Hydrodynamic fluctuations, broken symmetry, and correlation functions. – CRC Press, 2018.
7. Wang C.H. *Spectroscopy of Condensed Media*. Academic Press, Orlando 1985;
8. Hansen J.P. and McDonald I.R. *Theory of Simple Liquids, 2nd edu.* Academic Press, London 1986;
9. Balesku R. Equilibrium and Nanoquibrium statistical mechanics, Wiley.- New-York. – 1975.
10. Resibois P., Leener M. D. Classical kinetic theory of fluids //John Wiley & Sons 1977.

11. Berne B.J., Pecora R. In: *Dynamic Light Scattering, ed. Berne B.J.* John Wiley, N.Y. 1976;

12. Balucani U. and Zoppi M. *Dynamics of the Liquid State.* Clarendon Press, Oxford 1994.

13. Zubarev, D. N., Morozov, V. G., & Röpke, G. (1996). *Statistical Mechanics of Nonequilibrium Processes, Volume 1 (See 3527400834): Basic Concepts, Kinetic Theory* (Vol. 1). Wiley-VCH.

14. Statistical Mechanics of Nonequilibrium Processes. Vol. 2. Relaxation and Hydrodynamic Processes." *Statistical Mechanics of Nonequilibrium Processes, Wiley* (1997).

*15. Doi M., Edwards S.F.*, The Theory of Polymer Dynamic, Oxford: Clarendon Press, 1986.

*16. de Gennes P.G.*, Scaling Concepts in Polymer Physics, Ithaka : Cornell Univ. Press, 1970.

17. *Schweizer K.S.*// J. Chem. Phys., 1989, **V. 91**, P.5802.

18. Schweizer KS (1989) J Chem Phys,. 1989, V. 91: 5822

19. R Kimmich and N. Fatkullin, Adv. Polym. Sci. 170, 1 (2004).

20. M. A. Krutyeva, N. F. Fatkullin and R. Kimmich, Polymer Science, Ser. A, Vol. 47, No. 9, 2005, pp. 1022–1031.

21. N. F. Fatkullin, T. M. Shakirov, and N. A. Balakirev, Polymer Science, Ser. A, 2010, Vol. 52, No. 1, pp. 72–81.